%% file: main_arxiv.tex
\documentclass[11pt]{article}

\usepackage{setspace,graphicx,epstopdf,amsmath,amsfonts,amssymb,amsthm}
\usepackage{sty/versionPO}
\usepackage{marginnote,datetime,enumitem,subfigure,rotating,fancyvrb}
\usepackage{bm}
\usepackage{hyperref,float}
\usepackage{authblk}
\usepackage{hyperref}
\usepackage{booktabs}
\usepackage{ragged2e}
\usepackage{float}
\usepackage{threeparttable}
\usepackage{adjustbox}
\usepackage{csquotes}

\usdate

\excludeversion{notes}      %
\includeversion{links}      %

\iflinks{}{\hypersetup{draft=true}}

\ifnotes{%
\usepackage[margin=1in,paperwidth=10in,right=2.5in]{geometry}%
\usepackage[textwidth=1.4in,shadow,colorinlistoftodos]{todonotes}%
}{%
\usepackage[margin=1in]{geometry}%
\usepackage[disable]{todonotes}%
}

\makeatletter\let\chapter\@undefined\makeatother %

\usepackage{indentfirst}    %
\usepackage{footnote}      %
\usepackage{sty/jf}         %
\usepackage[labelfont=bf,labelsep=period]{caption}   %
\usepackage[backend=biber, style=apa, maxcitenames=1, natbib=true,doi=false]{biblatex}
\begin{document}
\setlist{noitemsep}  %
\onehalfspacing      %

\author[a]{Qiang Chen \thanks{\rm email: qiang2chen2@126.com}}
\author[b]{Tianyang Han \thanks{\rm email: tyhan@connect.hku.hk}}
\author[b]{Jin Li \thanks{\rm email: jli1@hku.hk}}
\author[b]{Ye Luo \thanks{\rm email: kurtluo@hku.hk}}
\author[c]{Zigan Wang \thanks{\rm email: wangzigan@sz.tsinghua.edu.cn}}
\author[b]{Yuxiao Wu \thanks{\rm email: wyx2000@connect.hku.hk}}
\author[c]{Xiaowei Zhang \thanks{\rm email: xiaoweiz@ust.hk}}
\author[b]{Tuo Zhou \thanks{\rm email: zhoutuo@connect.hku.hk}}
\affil[a]{School of Economics, Shandong University}
\affil[b]{HKU Business School, The University of Hong Kong}
\affil[c]{Tsinghua University}
\affil[d]{School of Engineering, Hong Kong University of Science and Technology}

\title{\Large \bf Can AI Master Econometrics? 
Evidence from Econometrics AI Agent on Expert-Level Tasks
}

\date{}

\maketitle
\thispagestyle{empty}

\centerline{\bf ABSTRACT}

    \noindent Can AI effectively perform complex econometric analysis traditionally requiring human expertise?  This paper evaluates AI agents' capability to master econometrics, focusing on empirical analysis performance. We develop ``MetricsAI'', an Econometrics AI Agent built on the open-source MetaGPT framework. This agent exhibits outstanding performance in: (1) planning econometric tasks strategically, (2) generating and executing code, (3) employing error-based reflection for improved robustness, and (4) allowing iterative refinement through multi-round conversations. We construct two datasets from academic coursework materials and published research papers to evaluate performance against real-world challenges. Comparative testing shows our domain-specialized AI agent significantly outperforms both benchmark large language models (LLMs) and general-purpose AI agents. 
    This work establishes a testbed for exploring AI's impact on social science research and enables cost-effective integration of domain expertise, making advanced econometric methods accessible to users with minimal coding skills. Furthermore, our AI agent enhances research reproducibility and offers promising pedagogical applications for econometrics teaching.

\bigskip



\clearpage

\section{Introduction}
\label{sec:Intro}

The rapid advancement of artificial intelligence (AI), particularly through large language models (LLMs) like ChatGPT (\cite{OpenAI:2024}), has sparked widespread discussion regarding its potential to enhance productivity, decision-making, and efficiency across diverse sectors. In recent years, these developments have gained notable traction in academia, especially in fields such as economics, business, and social sciences. Several studies (\cite{Li:2024}; \cite{Chakraborty:2025}) have already demonstrated that LLM-based tools can effectively perform tasks traditionally reserved for human participants, such as perceptual analysis and personnel selection, with remarkable accuracy and minimal professional intervention.

However, despite these promising advances, the adoption of AI for domain-specific analytical tasks remains limited, particularly in areas requiring complex econometric reasoning. Current LLM applications tend to focus on relatively simple, language-based tasks or the generation of basic numerical summaries from unstructured text. Scholars such as \textcite{Korinek:2023} and \textcite{Xu:2024} have highlighted persistent limitations in LLMs' ability to execute multi-step, logically constrained analytical workflows, especially when domain-specific methodological knowledge is required. \textcite{Ling:2024} further pointed out that fine-tuning LLMs for specialized fields like econometrics incurs substantial costs and struggles with maintaining pace with rapidly evolving academic frontiers.

From a broader economic perspective, the barriers to acquiring econometric and applied statistical skills present significant productivity losses and restrict the democratization of evidence-based research capabilities. In the absence of automation and AI support, performing rigorous empirical research typically demands years of specialized education, fluency in statistical programming, and deep familiarity with econometric theory and diagnostic conventions. This creates high entry barriers, disproportionately excluding students from under-resourced regions, non-English-speaking countries, and disciplines where quantitative literacy is not traditionally emphasized.

Economic theory predicts that lowering the cost of learning applied statistics and econometrics generates substantial productivity and welfare gains. In a simple production function model of knowledge accumulation, reducing the skill acquisition cost raises the equilibrium number of agents engaging in empirical analysis and policy-relevant research, thereby generating additional societal value. In particular, democratizing access to rigorous empirical tools is crucial for bridging global educational inequalities, especially in developing economies lacking experienced academic instructors in econometrics and applied quantitative research.

Building upon these motivations, this paper introduces MetricsAI (Econometrics AI Agent) — a specialized AI agent designed to perform applied econometric analysis via interactive, no-code workflows. Developed on the open-source MetaGPT framework, MetricsAI overcomes traditional LLM limitations by integrating a bespoke econometric tool library, a zero-shot learning framework, and a multi-round conversational workflow. This agent strategically plans econometric tasks, automatically selects appropriate econometric models based on data types and research objectives, generates and executes Python code, and iteratively refines results through error-based reflection and user feedback loops.

This innovation arrives at a time when AI agent frameworks are rapidly transforming industrial applications but remain under-explored in supporting academic research. Major technology leaders have emphasized the imminent expansion of agentic workflows. Andrew Ng (Stanford University), speaking at the AI Ascent 2024 Conference, predicted that "the set of tasks AI could do will expand dramatically this year because of agentic workflows." Similarly, Jensen Huang, CEO of Nvidia, projected at the 2024 Gartner IT Symposium/Xpo that his company would one day employ 50,000 people alongside over 100 million AI assistants. These endorsements reflect a growing consensus in both academia and industry that AI agents will redefine complex task automation in the coming years.

While AI agents such as Amazon’s Alexa+ already manage household devices through conversational interfaces, their academic counterparts remain scarce. Existing AI research tools either focus on unstructured data summarization or rely on static, fine-tuned models ill-equipped for dynamic, iterative research processes. Our paper directly addresses this gap by designing an AI agent tailored for the intricate demands of applied econometrics, social science, and business research.

Technically, MetricsAI represents a substantial advance over conventional LLM workflows. Most contemporary LLMs struggle to manage multi-step empirical workflows involving data preprocessing, model selection, estimation, inference, robustness checks, and diagnostic interpretation. MetricsAI resolves these issues through three core innovations:

1.	A custom econometric tool library implementing essential methods such as OLS, PanelOLS, IV-2SLS, DID, RDD, and Propensity Score techniques;
2.	A zero-shot learning framework allowing the agent to integrate new econometric procedures without costly LLM retraining;
3.	A multi-round interactive workflow enabling users to iteratively refine tasks through natural language queries, supported by contextual memory management and an error-based feedback mechanism.

To empirically validate MetricsAI’s performance, we conducted extensive tests using two real-world datasets: (i) doctoral-level econometrics coursework assignments and (ii) published empirical economics papers. The experimental results reveal substantial performance gains. While baseline LLMs such as GPT-4o achieved under 50\% success rates in complex tasks, and general-purpose AI agents achieved partial replication rates around 30\%, MetricsAI consistently attained directional replication rates exceeding 90\%, with perfect replication exceeding 50\% for coursework assignments and 27\% for published paper replications.

Beyond technical performance, this paper examines the broader economic, educational, and labor market implications of AI-driven econometric analysis tools. First, MetricsAI lowers the skill threshold for empirical social science research. By providing a no-code, interactive analysis environment, it allows undergraduate students and non-specialist professionals to replicate published econometric studies, a task traditionally reserved for doctoral students or seasoned researchers. This democratization effect is expected to expand the population of competent empirical researchers, improving the reach, readership, and citation potential of academic work.

Second, MetricsAI addresses barriers posed by language and technical fluency. The agent processes instructions in natural language, including non-English languages when necessary, and generates multilingual explanations of statistical outputs. This feature is particularly valuable for researchers and students in non-English-speaking regions, where access to quantitative economics training remains limited. By bridging these divides, MetricsAI contributes to educational equality and fosters the dissemination of evidence-based policy knowledge in under-resourced countries.

Third, the system promotes rapid adoption of state-of-the-art econometric methodologies. Empirical researchers frequently struggle to implement new techniques due to steep learning curves and sparse training materials. MetricsAI’s modular, zero-shot architecture allows newly published econometric methods to be incorporated into the tool library without requiring LLM retraining. Researchers can immediately access these methods via natural language queries, enhancing research productivity and reducing the prevalence of outdated or misapplied models in empirical studies.

Fourth, the agent contributes to improved research reproducibility. Replicability remains a significant concern in applied economics and business research (Mueller-Langer et al., 2019). By standardizing analytical workflows, automating diagnostic checks, and generating machine-readable results, MetricsAI facilitates rigorous, transparent, and easily audited empirical work. Its iterative, conversational interface also allows users to revise model specifications and robustness checks interactively, a capability largely absent in existing AI research tools.

Fifth, the deployment of MetricsAI has labor market implications for economics and business schools. As econometric literacy becomes increasingly accessible, universities are likely to expand their course offerings in applied statistics, business analytics, and causal inference. This expansion is expected to raise demand for economics instructors specializing in empirical research methods, potentially increasing employment opportunities in academic and professional training institutions. Moreover, enhanced research capacity in developing countries may stimulate local labor markets for empirical researchers and policy analysts, contributing to regional human capital development.

Finally, by dramatically reducing the technical and linguistic barriers to applied empirical analysis, MetricsAI paves the way for quantitative reasoning to become a basic educational competency akin to high-school mathematics. In the long run, it may help position data literacy and causal reasoning as foundational components of undergraduate education across disciplines, from economics and business to sociology and political science.

MetricsAI’s extensibility further enhances its long-term value. The system’s zero-shot learning framework allows new econometric techniques and domain-specific modules to be seamlessly integrated through Python functions and internal prompts, without altering the core LLM. This modularity ensures the agent’s continued relevance in a rapidly evolving research environment. Its architecture could also be transferred to other specialized fields, such as macroeconomics, finance, and quantitative public health, enabling the efficient development of AI agents equipped with field-specific analytic tools.

In sum, this paper makes three major contributions. First, we introduce a highly capable, cost-effective AI agent for applied econometric analysis. Second, we empirically demonstrate its superior performance relative to LLM baselines and general-purpose AI agents. Third, we offer an economic and educational analysis of MetricsAI’s potential impact on labor markets, higher education, research productivity, and global educational equality. The agent is open-source and available for public use at \url{https://github.com/HKU-Business-AI-Center/Econometrics-Agent}.

The remainder of the paper is organized as follows. Section \ref{sec:literature_review} reviews the emerging literature on LLM and AI agent applications in business, economics, and social sciences, positioning our contributions within this context. Section \ref{sec:Method} details the MetricsAI agent’s structural design, focusing on its domain-specific tool integration and task decomposition capabilities. Section \ref{sec:Evaluate} outlines our empirical evaluation methodology and presents comparative performance results against other LLM-based solutions. Section \ref{sec:case_studies} provides detailed case studies illustrating MetricsAI’s operation and advantages. Section \ref{sec:Conclusion} discusses the broader economic implications and future applications, concluding with remarks on its potential to reshape applied social science research.

\section{Literature Review}
\label{sec:literature_review}

Our paper relates to the growing literature on LLM applications in business, economics, and social sciences, where LLMs primarily serve as tools for analyzing unstructured textual data, replacing traditional natural language processing techniques. Research has demonstrated LLMs' fundamental content creation capabilities in various settings. \textcite{Stroube:2024} used GPT-4 to generate synthetic gender-neutral movie pitches to control for gender influence in quality evaluation. \textcite{Yoganarasimhan:2024} applied GPT-3.5 Turbo to calculate polarization scores for articles and topics in social media analysis. \textcite{Abraham:2024} utilized ChatGPT to generate ESG scores from PE firms' websites, improving upon dictionary-based measurements.
\textcite{Armstrong:2024} applied the GPT-3.5 Turbo to quantify the firm exposure in Securities Exchange Commission agencies but received insignificant enhancement.
\textcite{Niu:2024} verified the consistency of ``bag-of-words" technique by GPT-4 when measuring the effect of manufacturers’ service offerings on demand variability.
\textcite{Noailly:2024} utilized EnvP-BERT fine-tuned with environmental policies to capture climate-related indexes, which obtained resembling accuracy.
Apart from generating numeric data by analyzing the implication of the context, LLMs are additionally exerted in detecting the similarity by tokenizing context as vectors and afterward calculating the distances in different norms (\cite{Bursztyn:2022}; \cite{Graeber:2024}; \cite{Ahrens:2024}; \cite{Gorodnichenko:2024}; \cite{Curti:2023}; \cite{Balsmeier:2024}). 

Most existing research focuses on direct applications of LLMs' language capabilities, primarily using simple prompts to extract information from textual data. Due to LLMs' limited capabilities in specific domains like econometrics, their application in these areas remains restricted. Our paper advances this literature by endowing LLMs with domain knowledge through the MetricsAI's framework. This enables social science researchers to efficiently implement advanced econometric methodologies, expanding LLMs' utility beyond basic content generation.

Specifically, there have been recent studies discussing potential revolutionary roles of LLMs in econometrics. 
From a theoretical perspective, \textcite{Manning:2024} investigated LLMs' capability in social science research through automated structural causal model development and in silico simulations. \textcite{Ludwig:2025}  derived conditions under which LLMs can effectively conduct economic measurement prediction and estimation tasks. 
From an empirical perspective, \textcite{Han:2025} developed a multi-step prompt engineering method to validate LLMs' capability in identifying valid IVs for causal discovery. Our paper contributes directly to LLMs' empirical application in econometrics, focusing on automating researchers' complete hands-on workflow. Moreover, thanks to its extensible workflow and tool package design, our MetricsAI's framework supports smooth integration of fast-evolving AI-driven econometrics methodologies from academia, while serving as a platform for deploying these research outcomes in real-world applications.

Our second contribution relates to the literature exploring AI's economic impact and human-AI interaction. Previous research has extensively studied productivity effects. 
\textcite{Hui:2024} found that ChatGPT usage among freelancers reduces skill-based gaps in highly affected occupations, impacting both earnings and employment. \textcite{Brynjolfsson:2024} demonstrated heterogeneous productivity gains among customer service workers using LLM-based conversational guidance systems. Similar productivity effects were found by \textcite{Noy:2023} in professional writing and \textcite{Dell:2023} in consulting, where AI augmentation improved performance within AI-capable tasks but showed limitations beyond the AI's capability frontier.

These productivity shifts have broader implications, as \textcite{Meltzer:2024} discussed in analyzing potential job market changes across manufacturing and international trade, emphasizing the need for regulatory frameworks. 
Beyond productivity, Generative AI influences other economic behaviors. \textcite{Leib:2023} found that AI-generated dishonesty-promoting suggestions affect human honesty similarly to human advice, while \textcite{Doshi:2024} revealed that LLMs enhance individual creative writing but may reduce global novelty by channeling creativity in similar directions.

Research has also examined LLMs' decision-making patterns compared to humans. \textcite{Goli:2024} found LLMs prefer lower discount rates and earlier rewards in intertemporal choices, with effects moderated by language features. \textcite{Chen:2025} identified varying risk and certainty preferences across different LLM models in operations management decisions. Regarding occupational assessment, \textcite{Gmyrek:2024} found GPT-4 generally aligns with human judgments, except for overestimating digital economy roles and underestimating non-conventional occupations.

While these studies explore LLMs' economic impact across various dimensions, their research designs focus on LLMs' current capabilities and limitations. 
None examines the impact of domain-specific AI agents, which represent a more advanced application of LLMs with enhanced robustness and task-solving capabilities. Our paper addresses this gap. Through our agent design, we demonstrate how LLMs can transcend basic content generation to become effective task solvers and productivity enhancers. By studying how professionally-equipped AI agents impact practical users, we expand the scope of social science research into advanced AI applications.

Technically, our paper directly links to the concept of expert system, the computer program that emulates the behavior of a human expert within a well-defined domain of knowledge (\cite{Liebowitz:1995}), which is widely discussed since the last century (e.g., \cite{Tzafestas:1993}; \cite{Tan:2017}). The most traditional expert system is based on pre-defined rules, high-specified knowledge and various possible keywords, which strictly limit users' flexibility when interacting with the system to solve complex tasks. Another emerging branch applies neural network models and focuses more towards the learning capability of the system. Difficulty appears when facing insufficient training samples or requiring reasoning processes. Our MetricsAI enlightens a new direction for the future development of expert systems, combining the rapidly developing LLM techniques with the knowledge library and hence allowing the flexible functionality towards domain knowledge tasks.

Our paper makes another technical contribution addressing the challenge of enhancing LLMs' domain knowledge performance through a cost-effective solution. 
Current approaches primarily rely on expensive and time-consuming methods. These include fine-tuning models with curated domain-specific data, such as BloombergGPT (\cite{Wu:2023}) for finance and Med-PaLM 2 (\cite{Singhal:2023}) for medicine, or using lightweight fine-tuned LLMs to guide other LLMs (\cite{Yao:2023}). Alternative approaches focus on complex prompt engineering to generate domain knowledge content (\cite{Liu:2025}). However, without proper workflow and self-reflection mechanisms, these direct LLM applications remain vulnerable to hallucination issues that can significantly impact output quality. 

MetricsAI's structural design implements a zero-shot learning framework that ensures high task-solving performance at minimal cost. Our approach minimizes hallucinations through two key mechanisms. First, domain knowledge toolkit integration prevents hallucinations in code generation and execution, limiting potential errors to tool selection and calling procedures. Second, self-reflection processes further reduce hallucinations by validating execution outputs and implementing error corrections.
This design framework can be extended to other domains, particularly those with limited textual training samples or frequent knowledge updates, enabling efficient development of customized AI agents with domain-specific tools.

\section{Methodology}
\label{sec:Method}

Current LLM tools face two primary limitations: they struggle to deliver complete workflows and perform poorly in knowledge-dense domains. MetricsAI addresses both challenges through a customized AI Agent system and a zero-shot learning framework, enabling accurate and executable workflows without the substantial costs of LLM retraining.

\subsection{Overview of MetricsAI Structure}

LLM-based AI agents are autonomous systems that perceive instructions, reason about complex tasks, and execute actions to achieve specified goals through code execution and tool usage. 
While generic LLM agents can handle open-ended queries, they struggle in expert domains such as econometrics, where problem-solving requires adherence to established analytical steps and specialized methods. 
Recent AI agent frameworks demonstrate that incorporating structured domain knowledge significantly enhances performance. 
MetaGPT (Hong et al., \citeyear{MetaGPT:2024}a) encodes procedural knowledge as Standardized Operating Procedures within prompt sequences, using specialized sub-agents to verify intermediate results and decompose complex tasks. 
The Data Interpreter agent (Hong et al., \citeyear{datainterpreter:2024}b) uses a hierarchical task graph and iterative refinement modules for end-to-end data-science workflows.

Building on these advances, MetricsAI features a customized architecture specifically designed to embed econometric domain expertise into its core operations. This specialization moves beyond the capabilities of general-purpose agents or direct LLM interactions by incorporating specific structural enhancements tailored for econometric analysis.

\begin{figure}[ht]
    \centerline{\includegraphics[width=\textwidth]{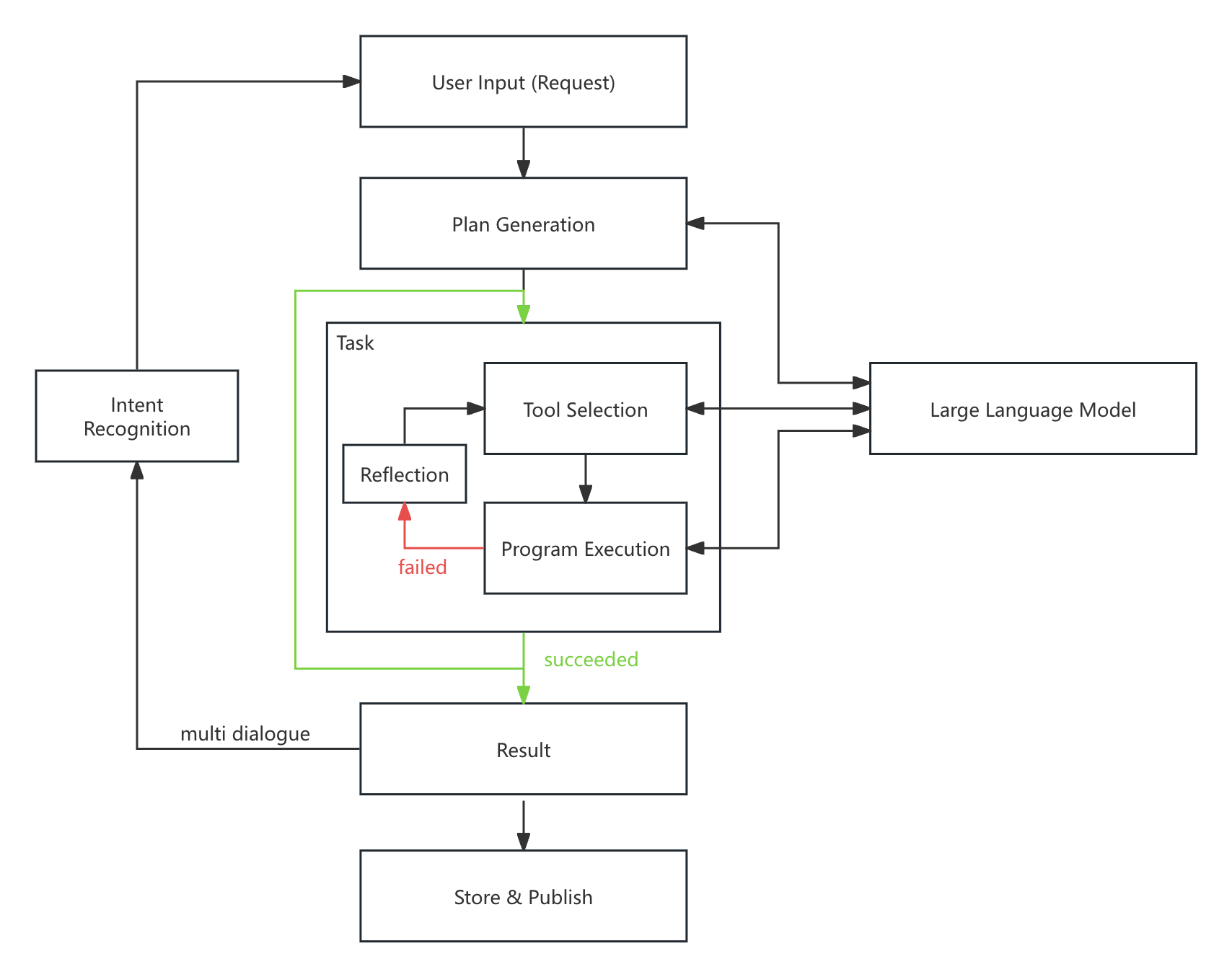}}
    \caption{{\bf Workflow of MetricsAI}}
    \label{fig:workflow_of_ai_agent}
\end{figure}

A cornerstone of this specialized structure lies in enhanced task decomposition. 
When presented with an econometric problem (\emph{User Input} in Figure \ref{fig:workflow_of_ai_agent}), the agent's \emph{Plan Generation} module follows predefined templates reflecting standard econometric research paths---such as causal inference strategies and time-series analysis workflows---rather than relying solely on generic planning. 
This approach decomposes complex requests into logically sequenced, econometrically meaningful sub-tasks. 
These sub-tasks are categorized using specific econometric actions like ``instrumental variable selection'' and ``difference-in-differences pre-trend check,'' enabling more precise execution than generic labels.

Crucially, executing these specialized sub-tasks requires more than the LLM's reasoning capabilities. 
Thus, another key structural enhancement is the integration of a unique Econometrics Tool Library. 
During the \emph{Tool Selection} and \emph{Program Execution} phases (Figure \ref{fig:workflow_of_ai_agent}), the agent accesses a bespoke set of Python functions implementing various econometric methods (e.g., OLS, PanelOLS, IV-2SLS, DID, RDD, propensity score methods), 
including options for robust standard errors and fixed effects.
As detailed in Section~\ref{subsec:custom}, these tools are described with internal prompts that enable the LLM core to understand their functionality and application context. 
This structured approach, where the agent selects and invokes validated tools,  minimizes LLM hallucination risk for complex algorithms and ensures adherence to econometric practices---a distinct advantage over general agents lacking such domain-specific implements.

Furthermore, recognizing empirical analysis as an iterative process, MetricsAI incorporates sophisticated multi-round conversational capabilities with intent recognition and memory. 
When users provide subsequent instructions (\emph{multi dialogue} loop in Figure~\ref{fig:workflow_of_ai_agent}), an \emph{Intent Recognition} mechanism assesses whether the new input relates to the ongoing analysis or initiates a new task. 
For continuing analyses, the agent retrieves context from its Memory---including current data state, generated code, existing plan, and environment variables within its sandboxed execution environment. 
It then updates the \emph{Plan Generation} phase, modifying or adding sub-tasks as needed, and resumes execution from the appropriate point. 
For new tasks, it initializes a fresh environment. 
This mechanism enables users to iteratively refine analyses, correct misunderstandings, or explore alternative specifications conversationally, ensuring outputs align with evolving research requirements.

These core structural enhancements---particularly the domain-tailored task decomposition and the unique econometrics toolkit---work in concert to ensure MetricsAI's effectiveness. 
The agent’s step-by-step reasoning is grounded in econometric best practices, while code execution relies on robust, pre-defined functions, leading to more reliable and accurate outcomes. 
This built-in domain expertise substantially reduces logical errors and methodological misapplications compared to general agents.

A key advantage of this design is its extensibility through a zero-shot prompting framework with tool use, allowing the agent to incorporate new econometric methods without LLM retraining.
Unlike the costly and often infeasible process of fine-tuning an LLM to keep pace with rapid academic advances in developing new techniques, our agent can be updated simply by adding new tool functions and descriptions to the prompt library. 
This modularity allows the agent’s knowledge base to expand alongside the field’s developments, 
making the integration of recently published procedures as straightforward as adding new modules. 
Such capability ensures the agent's state-of-the-art performance while potentially enabling the transfer of this structured approach to other specialized domains.

In short, by combining LLM-driven reasoning with domain-specific planning structures and a dedicated toolkit, MetricsAI achieves performance, robustness, accuracy, and adaptability beyond the capabilities of generic AI agents. 
The architecture executes complete econometric analyses end-to-end, delivering rigorous results while dynamically incorporating user feedback through its interactive workflow.

\subsection{Customization Towards Econometrics Tasks}\label{subsec:custom}

LLMs are generative pre-trained transformers that learn from massive amounts of textual data across various knowledge domains, enabling strong performance in general topics (\cite{Wang:2020}). 
However, achieving professional-level capability in specialized domains remains challenging, particularly in econometrics where two major limitations affect LLM fine-tuning.

First, econometric methods and algorithms rapidly evolve through academic journals, creating an expanding frontier of new knowledge. LLMs struggle to keep pace with these developments: frequent fine-tuning or retraining incurs prohibitive time and financial costs (\cite{Xia:2024}), while limited availability of training data---both textual explanations and code---for new methods constrains learning capabilities. 
Second, novel econometric methodologies often face slow adoption and knowledge diffusion among empirical researchers and practitioners, resulting in insufficient training samples for LLM learning and performance. Even for the emerging LLM-based tools with searching engines or retrieval-augmented generation (RAG) techniques, their capability in solving users' novel tasks will still be constrained when there are merely several academic papers available on the Internet.

The framework of MetricsAI achieves proper understanding and accurate application of econometric knowledge through a zero-shot learning approach. 
This is accomplished via an econometrics algorithm tool package designed specifically for LLM use rather than human users. 
Unlike traditional packages familiar to econometricians in Stata and R, our Python-scripted package incorporates system prompts engineered for LLM comprehension. 
The current tool package supports popular empirical econometric applications including general linear models, propensity score methods, IV-2SLS regression, static/staggered difference-in-differences, and sharp/fuzzy RDD. 
All econometric models and methodologies use \emph{function calling} format (\url{https://platform.openai.com/docs/guides/function-calling}), supporting flexible natural language inputs for econometric tasks.

More importantly, each function in the tool library includes a carefully designed internal prompt that introduces the background knowledge of the relevant econometric method and provides implementation guidance. 
These prompts define and explain hyper-parameters that enable flexible tool usage based on user requirements---for example, choosing between regular and robust standard errors in OLS. 
These system prompts enhance the LLM's understanding of different econometric tools' functionality, allowing accurate tool selection through reasoning when addressing user tasks.

An example of tool functions and internal prompts is provided in Figure~\ref{fig:prompt_example} in Appendix~\ref{sec:app1}. 
These tool packages are written in Python format to match MetricsAI's code execution environment. 
As previously introduced,
an LLM serves as the agent's core, understanding available tools, selecting appropriate ones for given tasks, and carefully extracting inputs from user commands, including datasets, econometric methodologies, and related hyper-parameters. 
Each tool features carefully designed functional arguments that meet researchers' needs while maintaining sufficient flexibility to handle diverse, customized econometric tasks.

The tool packages' functionality builds upon established Python packages for simpler econometric algorithms (like OLS and logistic regression), while complex algorithms lacking available packages (such as staggered DID event study and fuzzy RDD) are developed on our own following standard Python package formats. 
Each tool package includes a detailed description prompt---effectively a ``manual for LLMs''---that instructs the core LLM about the package's use. 
This design implements zero-shot learning \citep{Larochelle:2008,Lampert:2014}, 
allowing the LLM to learn directly from its environment and access relevant knowledge without training or fine-tuning costs.

An example of the complete tool-selection-and-implementation workflow is provided in Figure~\ref{fig:select_tool_example} in Appendix~\ref{sec:app1}. 
During initialization, the LLM processes each registered tool's internal prompt, summarizing four key aspects: target scenario, input requirements, output structure, and special requirements (if any). 
MetricsAI ranks available tools based on their target scenarios, selecting the highest-scoring tool for the current task. 
The LLM then generates code to ensure inputs meet tool package requirements while outputs fulfill user needs.

One might question whether LLM code generation capabilities could replace tools specifically designed for LLM use, suggesting that direct code generation might suffice given that Python and Stata package documentation is publicly available and likely included in LLM training data. 
However, limited training samples for specific econometric methods prevent LLMs from mastering domain knowledge and ensuring reliable performance. The test results in Sections \ref{sec:Evaluate} and \ref{sec:case_studies} demonstrate the critical importance of our tool package zero-shot learning framework.

Our design has several key advantages. 
First, it requires no econometrics-specific LLM tuning. General-domain LLMs (such as GPT-4o and LLaMA 3) perform effectively within MetricsAI, as demonstrated in our empirical tests using unmodified GPT-4o. 
Second, since all tool methods are fixed functions---although accepting flexible inputs---hallucination within econometric algorithms is eliminated. 
The LLM's role is limited to understanding tool application scenarios and selecting/calling appropriate tools, with any remaining errors readily caught by the code compiler.
Last but not least, the zero-shot learning framework enables simple integration of new algorithms and methodologies in Python function format, allowing continuous expansion of the econometric ``ecosystem'' and enhancement of the MetricsAI's capabilities.

\section{Empirical Tests}
\label{sec:Evaluate}

\subsection{Data Source and Summary Statistics}

We evaluate MetricsAI through two sets of inquiries. 
The first comprises 18 exercises from the coursework assignments of a doctoral-level course titled ``Applied Econometrics'' at the University of Hong Kong, with Python-generated standard solutions. 
These exercises cover OLS \& PanelOLS regression, propensity score matching, IV-2SLS regression, Difference-in-Differences (DID) analysis, and Regression Discontinuity Design. 
The second set consists of test datasets from randomly selected seminal articles in reputable journals, primarily accompanied by Stata-based replication packages.

\input{TablesAndFigures/4_A_test_summary}

\begin{figure}[ht]
    \centerline{\includegraphics[width=3in]{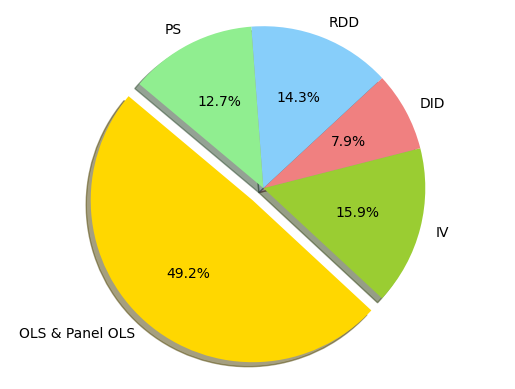}}
    \caption{{\bf Task Distribution (Econometric Methods)}}
    \label{fig:pie_distribution}
\end{figure}

Table \ref{tab:test_summary} and Figure \ref{fig:pie_distribution} show the distribution of econometric algorithms across both test datasets. OLS and PanelOLS emerge as the predominant methodologies in empirical research in business, economics, and social sciences, while advanced causal analysis techniques like IV and DID also make significant contributions. 
Our MetricsAI supports these algorithms, enabling automated research workflows with minimal development time and cost. All test datasets are available in the study's online appendix.

\subsection{Test Design}

Our empirical test adopts a structural prompt template to construct detailed questions. To demonstrate the MetricsAI's capabilities, we present results against control groups using standardized evaluation metrics.

\subsubsection{Prompt Structure}

While LLMs can process prompts of varying lengths, tones, attitudes, languages, and formats,
their performance may be affected by potential latent factors in these prompts \citep{He:2024}. 
To ensure consistent performance evaluation, we standardized the prompt format across all test questions. 
Each prompt must specify the data source, proposed econometric methodology, treatment variable, dependent variable, control variables, and customized requirements (such as data processing, fixed effects, robust standard errors, or methodology-specific settings). 
These components are integrated into a \emph{fixed} prompt structure. For illustration, consider this example from \textcite{Anderson:2008} using staggered DID event study to estimate safety belt legislation effects on traffic fatalities:

\begin{quotation}

    \textit{Please use the \textbf{Staggered DID Event Study} method to compute the effect of \textbf{``primary"} on \bm{$``log\_fatality\_rate"$}. \textbf{There is no control variable}.}
    
    \textit{Besides, you need to consider the following requirements: \textbf{divide states into two groups by the median value of population in 1982 and choose the high-population group. For the event study setting, set the see-back length as 4 and see-forward length as 3. Need to construct \bm{$log\_fatality\_rate$} as the dependent variable, by taking the direct logarithm of \bm{$fatality\_rate$}. Add state fixed effect, year fixed effect and cluster standard error by state. Output the result of coefficient towards \bm{$``Lag\_D1"$} term.}}
    
    \textit{You could load the corresponding data from \bm{$/home/data/DID\_sample\_data.dta$}.}
    
    \textit{At the end of the program, please print the coefficient, standard error and p-value of the effect in a json format like $\left\{ ``coefficient" \colon 1, ``standard\_error" \colon 2, ``p\text{-}value" \colon 0.5 \right\}$, and output the json string as a .json file.}
    
\end{quotation}

In this prompt template, boldface elements vary across questions to accommodate different econometric tasks, while the remaining structure stays fixed. 
For researchers familiar with econometric analysis, the workflow typically involves data exploration and cleaning, programming language selection, code implementation, and debugging. However, those without sufficient econometric experience face a more challenging process, often requiring extensive trial-and-error iterations.

\subsubsection{Test Procedure and Evaluation Metrics}

The generated prompts are fed directly to MetricsAI, which automatically executes Python code in a Jupyter kernel and iteratively updates based on error messages. As detailed in Section \ref{sec:Method}, the program terminates upon completion, storing results in JSON format. All scripts and outcomes are preserved for review, allowing users to interact with the AI agent through multi-round conversations for result revision and additional tasks.

Performance evaluation compares the generated coefficient, standard error, and p-value against standard solutions from coursework materials or original papers. We assess coefficient direction and $L_1$-norm distance (coefficients and standard errors in percentage, p-values in absolute distance). Two replication standards are defined: \emph{perfect replication} requires all three errors to be below $1\%$, while \emph{partial replication} requires coefficient and standard error gaps below $5\%$.

To demonstrate MetricsAI's capabilities, we establish three control groups:\footnote{To ensure clear performance differentiation across all the four groups, we consistently use GPT-4o as the base model for both direct response generation and agent empowerment. For the Stata-based tests, we manually collect metrics from summary tables, bypassing JSON file generation due to Stata's language constraints.}
\begin{enumerate}
    \item Direct GPT-4o Python code generation: We test the single-round code generation ability of GPT-4o by feeding it our structured prompts prefixed with ``use Python language to compute the following task'' and manually executing the generated code. This serves as a baseline for comparison.
    \item Direct GPT-4o Stata code generation: Similar to the first control, but testing LLM code generation in Stata, the most popular software in empirical research. Results are recorded in ``.smcl'' log files.
    \item General-purpose Data Interpreter agent (Hong et al., \citeyear{datainterpreter:2024}b): We test a standard AI agent for data analysis without econometric tool packages. This comparison highlights the advantages of our zero-shot learning framework and specialized tool package.

\end{enumerate}

\subsection{Test Result Summary}

Table~\ref{tab:assignment_result} summarizes the performance of the four groups on the coursework assignment problems. 
MetricsAI demonstrates superior performance with a 96.11\% directional replication rate and very low level of coefficient error (median value of errors only around 0.01\%). 
In contrast, both GPT-generated Python and Stata control groups show incorrect directions in over half of test cases. While the general AI Agent achieves a 77.78\% directional replication rate, its coefficient values frequently deviate significantly from true values, with the median value of coefficient errors reaching 11.06\%.

In addition, MetricsAI perfectly replicates around 58\% of cases, compared to approximately 10\% $\sim$ 27\% for control groups. Its robust econometric toolkit and systematic internal prompts yield higher replication rates for advanced algorithms and complex procedures. Furthermore, both AI agents - MetricsAI and general-purpose AI agent (Data Interpreter) - achieve almost perfect compilation rates through error-feedback mechanisms, while direct LLM generation groups compile successfully less than 36\% of the time.

\input{TablesAndFigures/4_C_assignment_summary}

Table \ref{tab:paper_result} presents results from the paper replication dataset. 
Given Stata's dominance in empirical research across business, economics, and social sciences, the Stata code generation group holds a natural advantage over Python-based approaches. Indeed, GPT-generated Stata code outperforms GPT-generated Python code across most evaluation metrics. 
However, MetricsAI achieves superior performance even in this Stata-favorable context, with a 88\% directional replication rate - 52\% higher than the Stata control group. Academic paper tasks present greater complexity than coursework, requiring detailed specifications, customized model structures, and various covariate adjustment methods. Nevertheless, MetricsAI consistently outperforms all control groups across key metrics: coefficient estimation error, standard error estimation error, p-value estimation error, and partial replication rates for different econometric algorithms and procedures. 
These results demonstrate the Agent's capability to handle complex econometric tasks autonomously while delivering reliable results.

\input{TablesAndFigures/4_C_paper_summary}

Despite substantially outperforming control groups, MetricsAI does show room for improvement. 
For example, its performance declines for complex econometric methods like DID and RDD compared to simpler approaches such as OLS and IV-2SLS. 
Similarly, results slightly deteriorate when moving from straightforward coursework problems to more sophisticated paper replication tasks. 
However, these limitations can be addressed through the AI agent's domain knowledge architecture---specifically by developing customized tools and enhancing prompt instructions to better support complex algorithms and detailed requirements.

\section{Case Study}
\label{sec:case_studies}

Following the empirical results presented in Section \ref{sec:Evaluate}, we demonstrate the MetricsAI's problem-solving approach through a comprehensive case study, comparing its procedures with the three control groups. 
The case study is drawn from \textcite{Almond:2005} with the theme of estimating the effect of maternal smoking during pregnancy on infant weights applying propensity score-related approaches.

The first component is derived from the doctoral-level econometrics coursework materials and employs a propensity score-based regression adjustment method to conduct the analysis. This method was first introduced in \textcite{ROSENBAUM:1983} and discussed by economists since \textcite{Card:1988}. With its advantages in maintaining sample completeness and supporting flexibility in non-linear \& interaction effects, as well as its model simplicity, this method is commonly used as the first-step analysis among studies with propensity score-based methods. The question prompt is as follows:

\begin{quotation}

    Please use the \textbf{propensity score regression} method to compute the effect of \textbf{tobacco} on \textbf{dbrwt}. You also need to control the following control variables: \textbf{rectype, csex, dmar, pldel3, pre4000, preterm, alcohol, dmage, demduc, dlivord, monpre, nprevist, dplural, birattnd, cntocpop, ormoth, mrace3, adequacy, delmeth5}.
    
    Besides, you need to consider the following requirements: \textbf{birattnd, cntocpop, ormoth, mrace3, adequacy, delmeth5 are multi-class categorical variables. Trim the samples with the highest 10\% score and the lowest 10\% score} 
    
    You could load the corresponding data from \bm{$/home/data/ps1\_24S\_cleaned.dta$}.
    
    At the end of the program, please print the coefficient, standard error and p-value of the effect in a json format like $\left\{ ``coefficient" \colon 1, ``standard\_error" \colon 2, ``p\text{-}value" \colon 0.5 \right\}$, and output the json string as a .json file.
    
\end{quotation}

The empirical research process begins with data pre-processing and matching, including data file merging, variable selection (dependent, independent, and control variables), and handling missing values. 
The analysis requires:
\begin{enumerate}
    \item Constructing propensity scores using logistic or probit models (given the binary treatment variable \texttt{tobacco}). 
    \item Trimming samples to exclude extreme propensity scores (e.g., below $0.1$ or above $0.9$) to ensure sufficient overlap between treatment and control groups. 
    \item Running OLS regression with \texttt{dbrwt} as the dependent variable and both \texttt{tobacco} and the propensity score as independent variables. 
\end{enumerate}
Using the provided dataset, the standard solution yields an ATE of $-207.7272$ with standard error $5.508$. 
One may also add the covariates into the second-stage OLS regression, which would produce an alternative ATE of $-212.9892$ with standard error $5.071$.

In the following, we compare MetricsAI with the three control groups in terms of problem-solving procedures, highlighting common LLM tool limitations and demonstrating how MetricsAI addresses them.
All code generation and execution records are available online and can be accessed via \url{https://huggingface.co/datasets/troyhan/econometric_ai_agent_testset/tree/main}.

\subsection{Python Code Generation: LLM and AI Agents}

Direct code generation by LLMs appears to be a straightforward solution given the clear requirements and simple knowledge base. 
However, LLM hallucinations frequently disrupt code generation, where even minor errors can cause program termination without any meaningful results. 
The GPT-generated Python script demonstrates this vulnerability. 
As propensity score regression adjustment lacks built-in Python implementation, LLMs must generate complete data processing and regression procedures. 
Without self-correction capabilities, hallucinations occur in two areas, causing coding errors and execution failure:

\begin{itemize}
    \item Syntax Error: After the ``pd.get\_dummies'' operation, all variables affected are transformed into new columns with new labels (Figure~\ref{fig:cs1_1} in Appendix~\ref{sec:app2}). 
    LLMs fail to detect this label change, during code generation, causing execution termination.
    
    \item Logic Error: During categorical variable preparation, GPT incorrectly applies dummy variable transformation to all covariates, including continuous variables, rather than only categorical predictors (Figure~\ref{fig:cs1_2} in  Appendix~\ref{sec:app2}). 
    This deviation violates problem requirements, 
    creating an oversized right-hand-side dataset (exceeding $112313 \times 112313$)   and triggering errors.
\end{itemize}

Unlike direct LLM code generation, 
the general-purpose AI agent avoids coding errors through its reflection capabilities.
However, Figure \ref{fig:cs1_3} in Appendix~\ref{sec:app2} reveals a \emph{behavioral bias} of the agent. 
Since logistic regression, the first step of this analysis, is common in machine learning,
the agent follows machine learning conventions by splitting data into training and test sets. This approach conflicts with standard econometric practice, which emphasizes empirical explanation rather than prediction, making such splitting unnecessary. 
The resulting estimates deviate significantly from the correct answer, yielding an ATE of $-103.7577$ with standard error $15.8834$.

Far from coincidental, this behavioral bias reflects a systemic issue stemming from the substantially greater accessibility of machine learning materials compared to econometric code resources.
MetricsAI addresses this bias through two mechanisms: detailed internal prompts that enforce econometric standards, and a specialized tool library that ensures adherence to econometric conventions.

\subsection{Stata Code Generated by LLM}

Since Stata provides numerous built-in data pre-processing methods and econometric algorithms, LLMs generate Stata code more effectively than Python code for econometric analysis. Moreover, Stata's specialized focus on econometrics, rather than machine learning, means its scripts in training samples naturally guide LLMs toward standard econometric approaches.
However, Stata's function library lacks a built-in function specifically for propensity score regression adjustment. 
After obtaining propensity scores, the optimal approach is to directly apply OLS for final estimation. 
Instead, as Figure~\ref{fig:cs1_4} in Appendix~\ref{sec:app2} shows, the hallucination issue arises, leading the LLM to mistakenly select propensity score matching for the second-step estimation. 
Although the results closely match the standard answer (ATE: $-218.9029$, standard error: $7.5627$), 
this methodological deviation render these estimates unreliable.

The \emph{knowledge mismatch error} is another aspect of the hallucination issue that requires careful attention. 
For tasks related to propensity score, although OLS regression adjustment is simpler, 
LLMs tend to select the more popular and well-documented propensity score matching method, following their tendency to generate the most probable tokens.

In contrast, MetricsAI has two advantages that prevent this error: First, its tool library supports user-customized functions tailored to specific needs, including less common methods. 
For this task specifically, the Agent is equipped with both propensity score regression adjustment functionality and various other propensity score-based econometric methods. Second, the selection of econometric algorithms (and their corresponding tool functions) does not rely on LLM generation capabilities. 
Instead, the tool recommendation procedure introduced in Section~\ref{sec:Method} first identifies the tool that best matches the task requirements. 
Only after that does LLM generate corresponding codes to correctly apply the tool and finish the task.

\subsection{MetricsAI Operation Record}

MetricsAI decomposes this task into three steps. The first two steps---``Load and preprocess the dataset from a specified file path'' and ``Perform exploratory data analysis on the dataset''---follow standard empirical econometric paradigms under internal prompt instructions. 
These steps complete essential data preparation, including categorical variable one-hot encoding and data cleaning. 
Furthermore, as Figure~\ref{fig:cs1_5} in Appendix~\ref{sec:app2} shows, MetricsAI's precise workflow requirements prevent the coding errors observed in control groups.

For the final step, ``Apply propensity score regression, controlling for specified variables and trimming samples,'' MetricsAI utilizes built-in tool functions to construct propensity scores and conduct regression adjustment analysis, as shown in Figure~\ref{fig:cs1_6} in Appendix~\ref{sec:app2}. 
After tool selection, the LLM focuses solely on preparing correct inputs to meet task requirements. 
This approach significantly reduces both the \emph{knowledge mismatch error} from LLM hallucination and the \emph{behavioral bias} through standardized tool functions. 
The final estimates (ATE: $-207.8559$, standard error: $5.4845$) closely match the standard answer, validating the results.

\subsection{More Illustration Towards Knowledge Hallucination Issue}

Beyond the discussions in \textcite{Almond:2005}, 
subsequent studies have further explored propensity score methods for treatment effect identification. 
Take \textcite{Cattaneo:2010} as an example. 
Based on their analytical frameworks, we formulate another propensity score matching task, which embeds the more commonly-used propensity score-based methodology, with the following instruction prompt:

\begin{quotation}

    Please use the \textbf{propensity score matching} method to compute the ATE of \textbf{tobacco} on \textbf{dbrwt}. You also need to control the following control variables: \textbf{mmarried, mage, mage2, fbaby, medu}.
    
    Besides, you need to consider the following requirements: \textbf{need to construct mage2 by taking the squared value of mage; use one-to-one matching; mbsmoke, mmarried, fbaby are dummy variables, and other variables are numerical variables} 
    
    You could load the corresponding data from \bm{$/home/data/cattaneo.dta$}.
    
    At the end of the program, please print the estimated ATE in a json format like $\left\{``ATE" \colon 0.2\right\}$, and output the json string as a .json file.
    
\end{quotation}

The manual solution procedure follows similar initial steps as the propensity score regression adjustment method. 
For ATE estimation, a nearest-neighbor matching method is applied to both treatment and control groups using estimated propensity scores. The final estimate is derived from the mean difference between groups across all included entities. Under the specified settings, this procedure yields an ATE estimate of $-210.9683$.

Since propensity score matching is a built-in Stata function, GPT-generated Stata code provides succinct solutions. However, for Python implementations, where no popular packages exist for this method, the generated code must implement the method step-by-step. Here, the \emph{knowledge hallucination issue} emerges consistently in our tests: as Figure~\ref{fig:cs1_7} in Appendix~\ref{sec:app2} shows, GPT models claim to calculate ATE while actually producing ATET (Average Treatment Effect on the Treated) estimates. 
This error stems from LLMs' domain knowledge limitations. While general-purpose AI agents can use internal prompts to guide LLM behavior, such guidance is not sufficient and cannot ensure comprehensive knowledge of every domain-specific detail.

This example therefore provides direct evidence for the advantage of MetricsAI's tool library design. 
Figure~\ref{fig:cs1_8} in Appendix~\ref{sec:app2} demonstrates how the core LLM, supported by a well-established knowledge and tool library, avoids knowledge hallucination under MetricsAI's framework. 
Rather than solely relying on LLMs' content generation ability to perform domain knowledge-driven tasks---which are often complicated, in-depth, and sometimes uncommon and ambiguous---the tool library greatly simplifies LLMs' generation and decision processes while guaranteeing domain knowledge capability. Beyond this straightforward case example and simple methodology, the tool library's robustness, flexibility, and extensibility ensure the elimination of knowledge hallucination across more complex tasks and algorithms.

\section{Conclusions}
\label{sec:Conclusion}

The AI era has brought significant productivity gains across various domains, yet challenges persist in fields requiring deep domain expertise. 
We introduce MetricsAI, a LLM-driven specialized system that automates econometric analysis. 
Through a carefully customized agent structure, it executes complete econometric analyses while dynamically adapting to user feedback. 
The agent incorporates a simple yet robust zero-shot learning framework that enables continuous functionality expansion, both within econometrics and across other knowledge domains. 
We demonstrate the agent's superior performance through empirical testing on coursework assignments and economics papers, complemented by in-depth case study comparisons against standard LLMs and general-purpose AI agents.

MetricsAI's excellent performance demonstrates its significant potential for future applications. Beyond reducing learning barriers for econometrics students, it provides academic researchers and industry practitioners with efficient tools for research tasks. As leading academic journals increasingly require original data and replicable procedures, manual paper proofreading has become time-consuming and challenging. 
MetricsAI can serve as an AI-driven digital referee for empirical research papers, significantly boosting proofreading efficiency (\cite{Mueller:2019}) while ensuring content quality and validity.

Our work's extensibility manifests in two key dimensions. First, MetricsAI's capabilities can be expanded by incorporating additional tool packages. For instance, addressing growing concerns about p-hacking in empirical research in business, economics, and social sciences, we developed a tool package using inverse optimization to analyze potential decision procedures regarding empirical test settings within given optimization spaces and datasets. 
This cost-effective addition helps detect potential p-hacking risks in empirical research, offering an efficient alternative to manual review.

Second, our zero-shot learning framework can extend to other knowledge-intensive domains. Through targeted instruction prompt refinement and domain-specific tool package development, this one-off setup enables new AI agents to deliver consistent, high-quality performance in their respective domains. This extensibility opens possibilities for future applications in areas such as quantitative investing and macroeconomics.

\linespread{1.3}\selectfont
\printbibliography

\clearpage

\appendix
\counterwithin{figure}{section}

\section{Tool Design, Tool Selection and Internal Prompt}
\label{sec:app1}

\begin{figure}[!htb]
    \centerline{\includegraphics[width=6.5in]{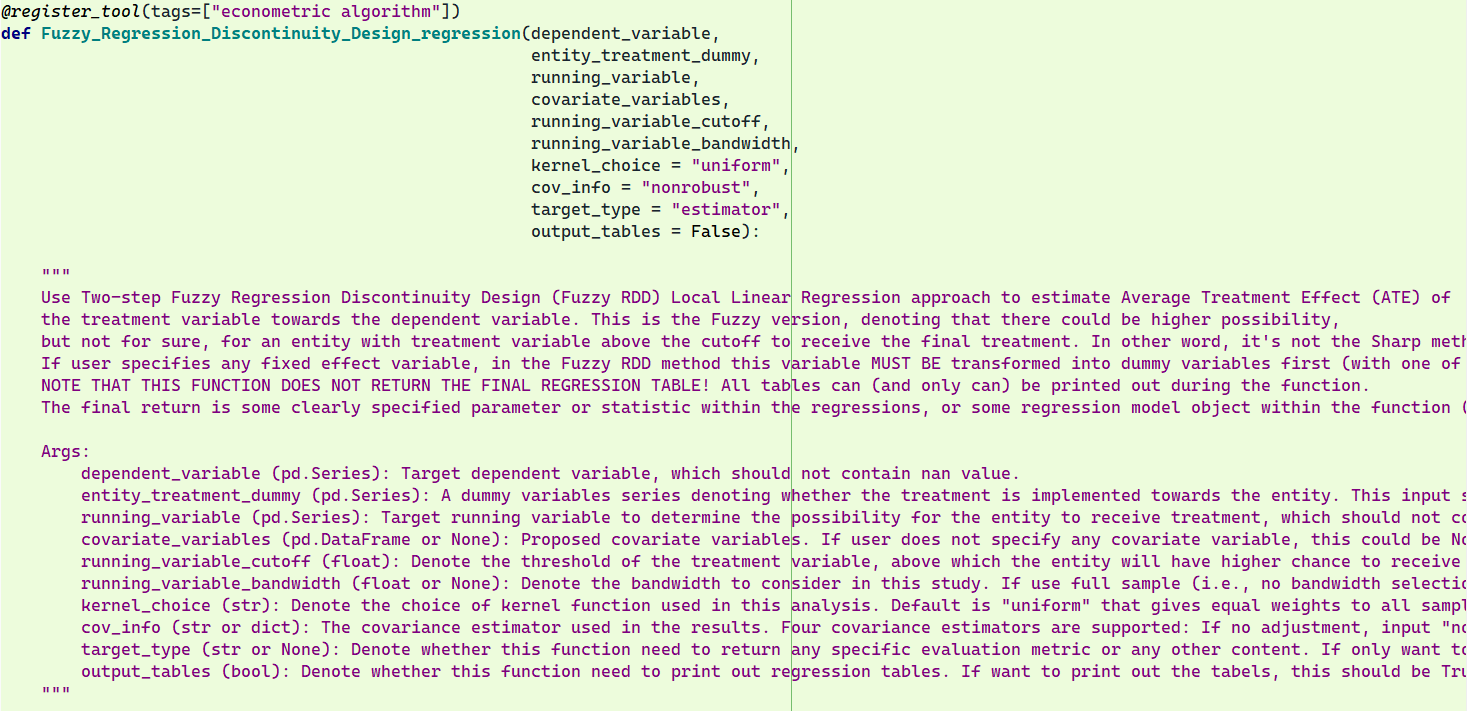}}
    \caption{{\bf An Example of Econometrics Tool and Internal Prompt}\\}
    \label{fig:prompt_example}
\end{figure}

\begin{figure}[!htb]
    \centerline{\includegraphics[width=6.5in]{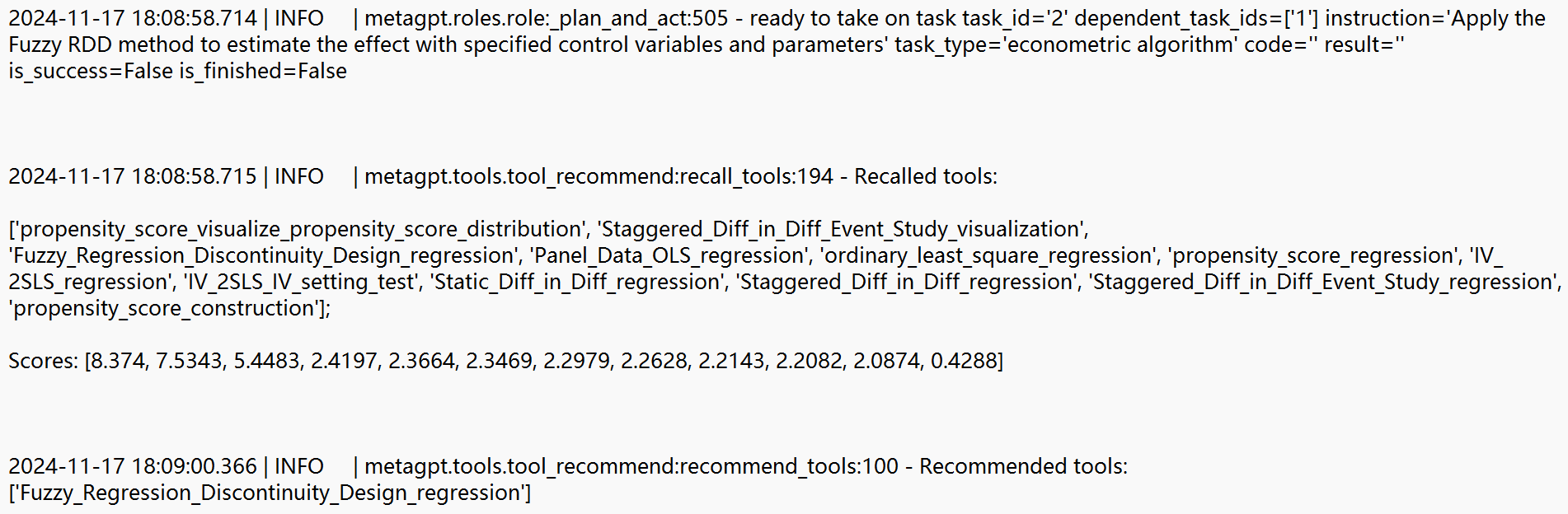}}
    \caption{{\bf An Example of Econometrics Tool Selection Procedure}\\}
    \label{fig:select_tool_example}
\end{figure}

\clearpage

\section{Case Study Response Examples}
\label{sec:app2}

\begin{figure}[!htb]
    \centerline{\includegraphics[width=6.5in]{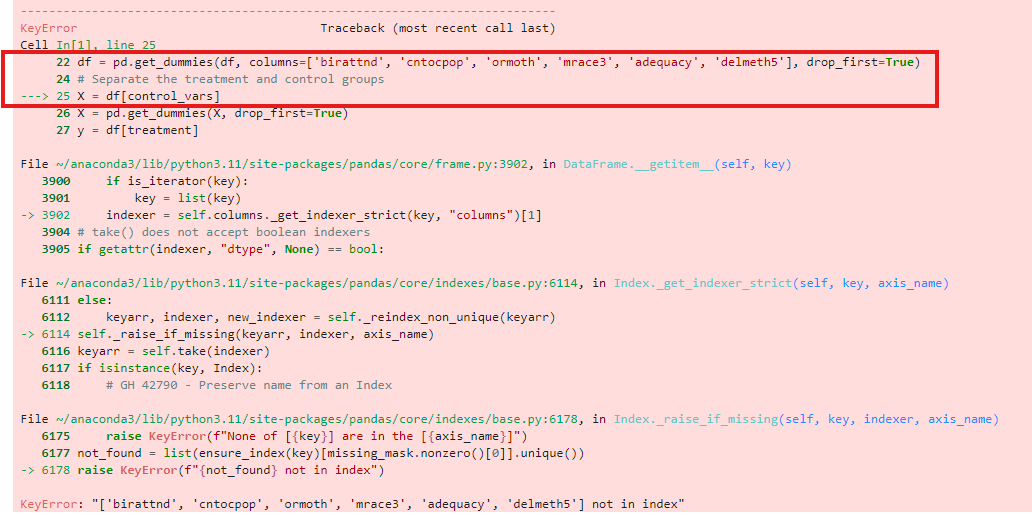}}
    \caption{{\bf Case Study --- GPT(py) Hallucination Problem 1}\\}
    \label{fig:cs1_1}
\end{figure}

\begin{figure}[!htb]
    \centerline{\includegraphics[width=6.5in]{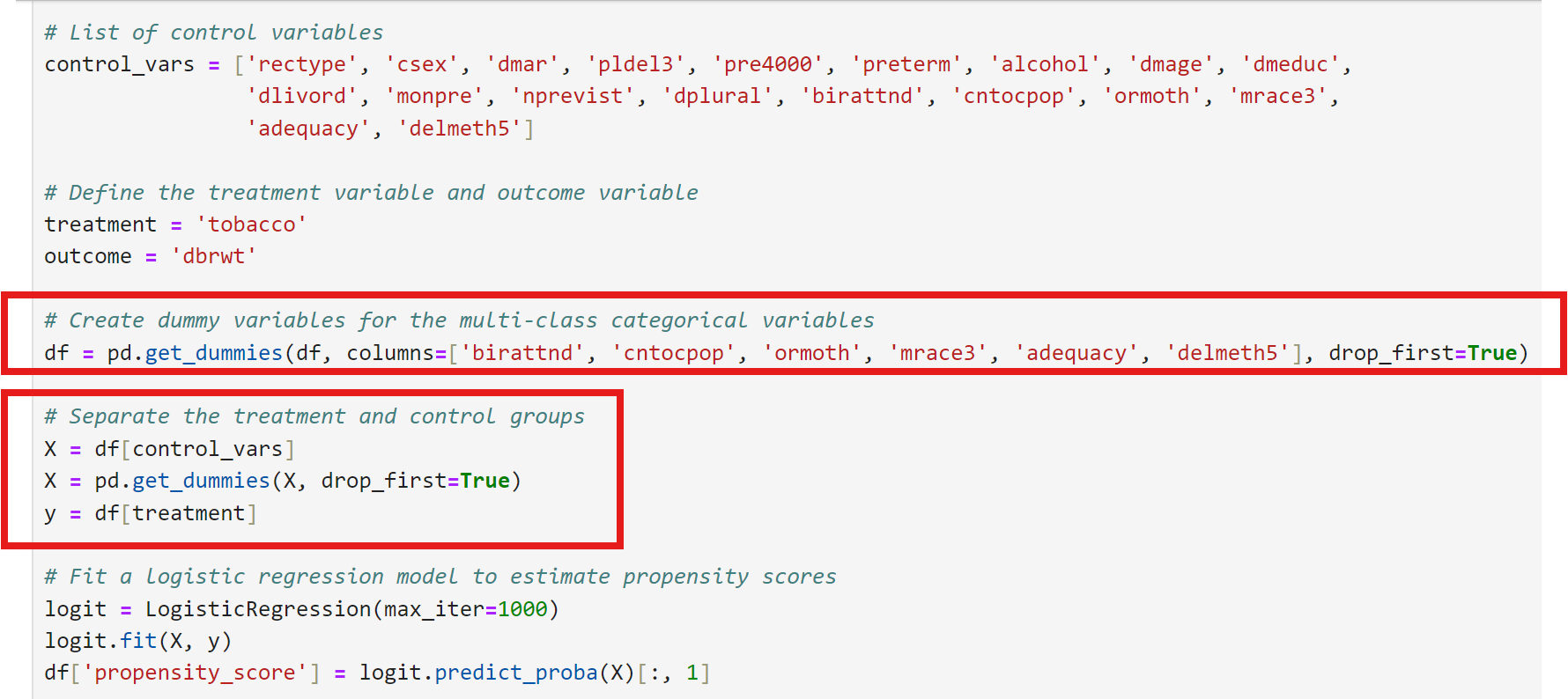}}
    \caption{{\bf Case Study --- GPT(py) Hallucination Problem 2}\\}
    \label{fig:cs1_2}
\end{figure}

\begin{figure}[!htb]
    \centerline{\includegraphics[width=6.5in]{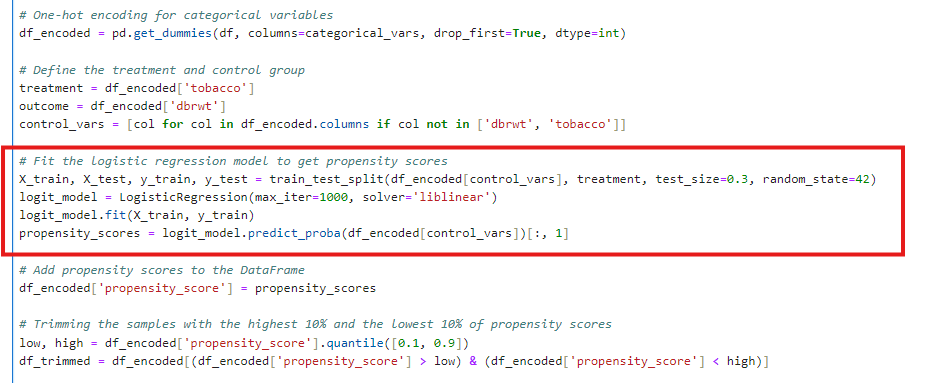}}
    \caption{{\bf Case Study --- General AI Agent Hallucination Problem}\\}
    \label{fig:cs1_3}
\end{figure}

\begin{figure}[!htb]
    \centerline{\includegraphics[width=6.5in]{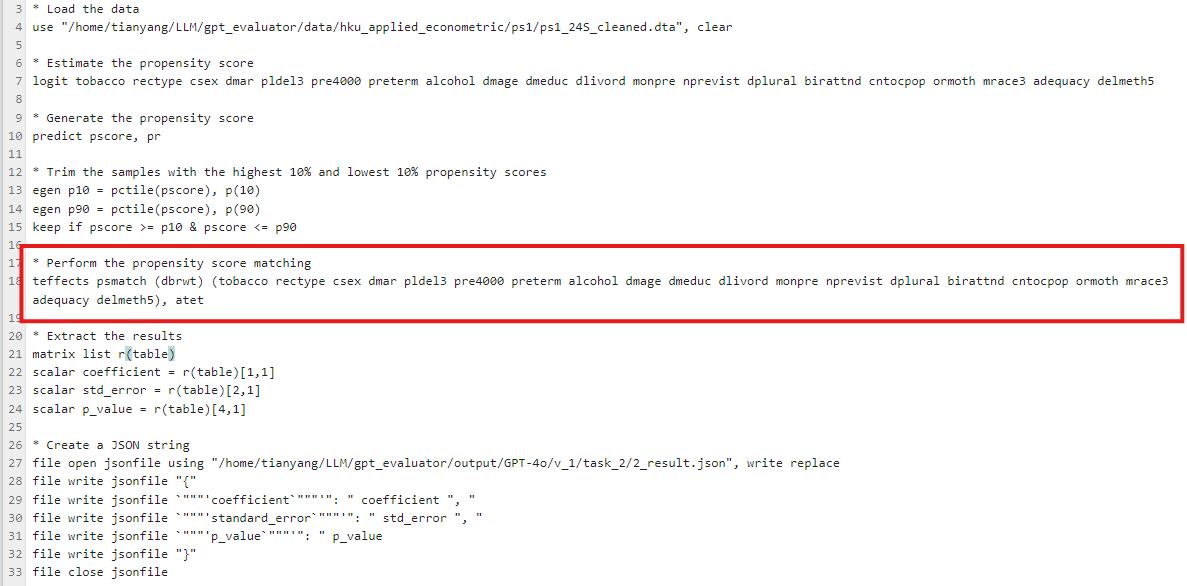}}
    \caption{{\bf Case Study --- GPT(stata) Hallucination Problem}\\}
    \label{fig:cs1_4}
\end{figure}

\begin{figure}[!htb]
    \centerline{\includegraphics[width=6in]{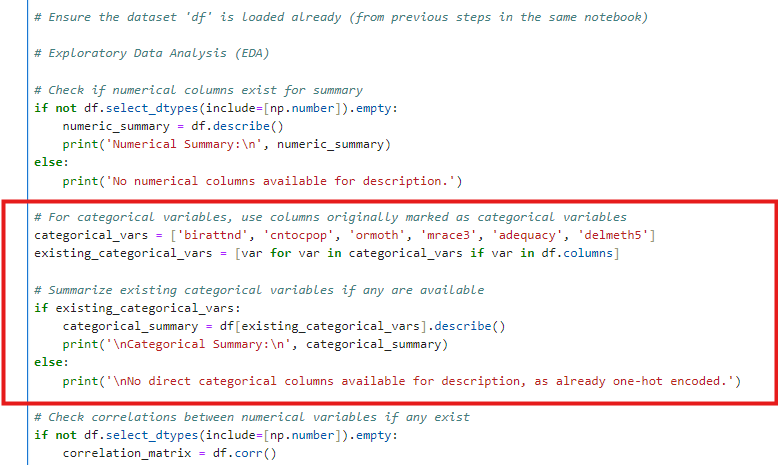}}
    \caption{{\bf Case Study --- MetricsAI Operation Record 1}\\}
    \label{fig:cs1_5}
\end{figure}

\begin{figure}[!htb]
    \centerline{\includegraphics[width=6in]{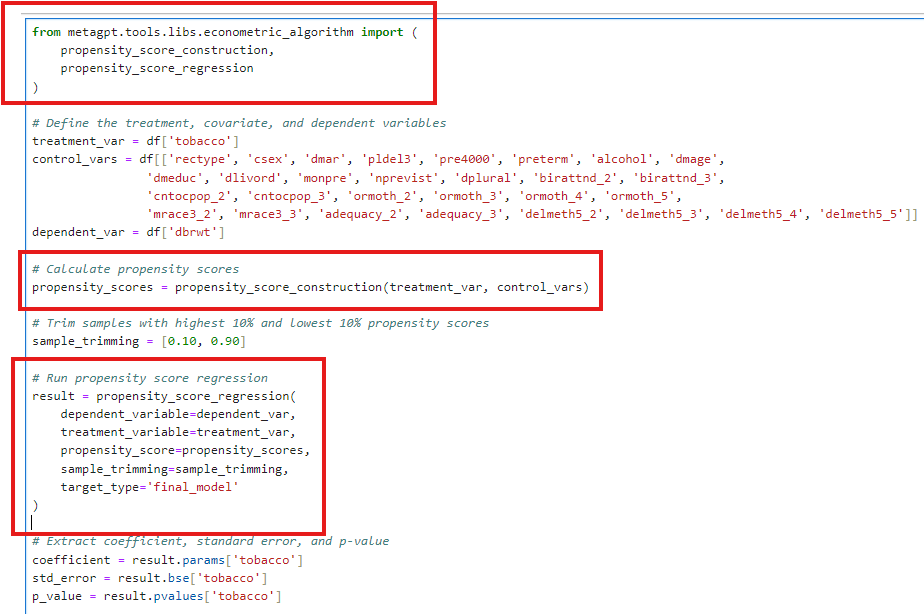}}
    \caption{{\bf Case Study --- MetricsAI Operation Record 2}\\}
    \label{fig:cs1_6}
\end{figure}

\begin{figure}[!htb]
    \centerline{\includegraphics[width=5in]{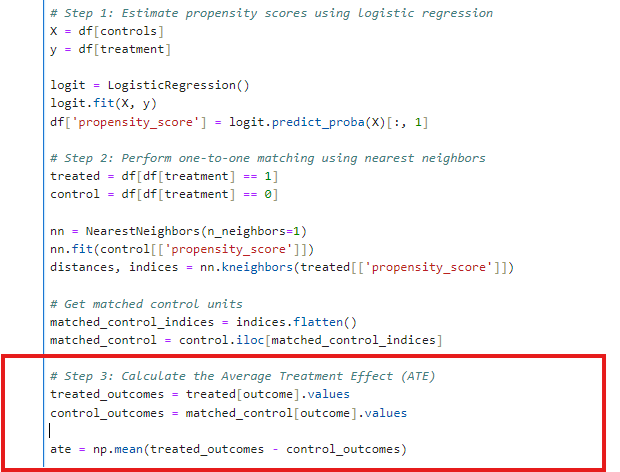}}
    \caption{{\bf Case Study --- Knowledge Hallucination Example}\\}
    \label{fig:cs1_7}
\end{figure}

\begin{figure}[!htb]
    \centerline{\includegraphics[width=7in]{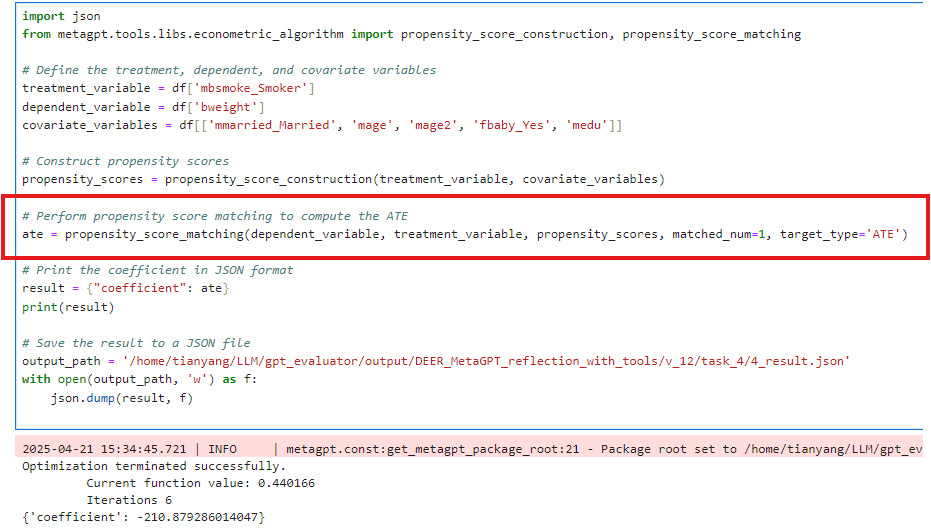}}
    \caption{{\bf Case Study --- MetricsAI Knowledge Hallucination Solution}\\}
    \label{fig:cs1_8}
\end{figure}

\clearpage

\end{document}

%% file: TablesAndFigures/4_A_test_summary.tex
\begin{table}[!htb]

\centering
\begin{adjustbox}{width=0.8\textwidth}
\scriptsize 
\begin{tabular}{lll}
 & \textbf{Assignments} & \textbf{Published Papers}\\ %
\toprule
\textbf{Number of Samples (Percentage)} \\
\cmidrule(lr){1-1}
OLS \& Panel OLS                     & 8 (44.4\%) & 23 (51.1\%) \\
Propensity Score                     & 2 (11.1\%) & 6 (13.3\%)  \\
Instrument Variable (IV)             & 2 (11.1\%) & 8 (17.8\%) \\
Difference-in-Differences (DID)      & 2 (11.1\%) & 3 (6.7\%) \\
Regression Discontinuity Design (RDD) & 4 (22.2\%) & 5 (11.1\%)  \\
\midrule
Data Processing               & 13 (72.2\%) & 23 (51.1\%)\\
Covariate Adjustment         & 14 (77.8\%) & 29 (64.4\%)\\
Fixed Effects                 & 4  (22.2\%) & 24 (53.3\%)\\
\midrule
Total                           & 18 &  45 \\
\bottomrule
\footnotesize
\end{tabular}

\end{adjustbox}
\caption{{\bf . Testset Summary Statistics}}
\label{tab:test_summary}
\end{table}

%% file: TablesAndFigures/4_C_assignment_summary.tex
\begin{table}[!htb]
\centering
\begin{adjustbox}{width=1\textwidth}
\scriptsize 
\begin{tabular}{lllll}
 & \textbf{GPT(Py)} & \textbf{GPT(Stata)} & \textbf{General AI Agent} & \textbf{MetricsAI}\\ %
\toprule
\textbf{Overall Performance} \\
\cmidrule(lr){1-1}
Compilation Success   & 35.56\% & 23.33\% & 96.67\% & 98.33\% \\
Perfect Replication   & 12.22\% & 10\%    & 26.67\% & 57.78\% \\
Partial Replication  & 16.67\%  & 11.11\% & 32.22\% & 66.11\% \\

\midrule
\textbf{Coefficient} \\
\cmidrule(lr){1-1}
Correct Direction          & 35.56\% & 21.11\% & 77.78\% & 96.11\%  \\
Coefficient Median Error   & 0.99\%  & 0.04\%  & 11.06\% & 0.01\% \\
Coefficient Error $<$ 1\%  & 17.78\% & 11.11\% & 16.67\%  & 68.33\% \\ 
Coefficient Error $<$ 10\% & 30\%    & 15.56\% & 38.89\%  & 78.33\% \\ 

\midrule
\textbf{Standard Error} \\
\cmidrule(lr){1-1}
Standard Error Median Error   & 2.14\% & 2.12\% & 8.01\% & 0.28\% \\
Standard Error Error $<$ 1\%  & 13.75\% & 8.75\% & 25\% & 55.62\% \\ 
Standard Error Error $<$ 10\% & 32.5\% & 17.5\% & 43.75\% & 80\% \\ 

\midrule
\textbf{P-value \& Significant Level} \\
\cmidrule(lr){1-1}
P-value Average abs Error         & .0197  & .0306   & .0584 & .0243 \\
P-value Median abs Error          & 0      & .0004   & 0     & 0 \\
P-value abs Error $<$ 1\%         & 37.5\% & 15\%    & 68.75\% & 89.38\% \\ 
P-value abs Error $<$ 10\%        & 37.5\% & 16.25\% & 68.75\% & 90.62\% \\ 
Significant Level Correctness     & 37.5\% & 17.5\%  & 68.75\% & 90\% \\ 
Significant Level Error == 1  & 0      & 0      & 0      & 2.5\% \\ 
Significant Level Error == 2  & 1.25\% & 0      & 6.25\% & 0 \\ 
Significant Level Error == 3  & 0      & 1.25\% & 6.75\% & 3.12\% \\ 

\midrule
\textbf{Partial Replication based on Task Type}\\
\cmidrule(lr){1-1}
OLS, Panel OLS                    & 32.5\%   & 7.5\%  & 37.5\% & 51.25\% \\
Propensity Score Regression       & 0      & 0     & 0      &  45\% \\
Instrument Variable (IV)          & 20\%   & 60\%  & 50\%   & 100\% \\
Differences in Differences (DID)  & 0      & 0     & 50\%   &  75\% \\
Regression Discontinuity Designs (RDD)  & 0 & 5\%  & 0      & 85\% \\
\cmidrule(lr){1-5}
Data Processing               & 16.92\% & 15.38\% & 30.77\% & 77.69\%\\
Covariate Adjustment         & 14.29\% & 14.29\% & 28.57\% & 65\% \\
Fixed Effects                 & 30\%    & 0       & 75\%    & 47.5\% \\
\bottomrule
Number of Tasks    & 90 & 90 & 180 & 180
\end{tabular}

\end{adjustbox}
\caption{{\bf . Assignment Testset Performance}}
\label{tab:assignment_result}
\end{table}

%% file: TablesAndFigures/4_C_paper_summary.tex
\begin{table}[!htb]
\centering
\begin{adjustbox}{width=1\textwidth}

\scriptsize 
\begin{tabular}{lllll}
 & \textbf{GPT(Py)} & \textbf{GPT(Stata)} & \textbf{General AI Agent} & \textbf{MetricsAI}\\ %
\toprule
\textbf{Overall Performance} \\
\cmidrule(lr){1-1}
Compilation Success   & 28.89\% & 38.89\% & 88.44\% & 96\% \\
Perfect Replication   & 8.44\%  & 17.78\% & 19.78\% & 32.67\% \\
Partial Replication   & 13.78\% & 22.22\%  & 36.22\% & 42\%  \\

\midrule
\textbf{Coefficient} \\
\cmidrule(lr){1-1}
Correct Direction          & 24.44\%    & 36.11\% & 74.67\% & 87.78\%  \\
Coefficient Median Error   & 0.86\%     & 0.00\%  & 3.24\%  & 0.16\% \\
Coefficient Error $<$ 1\%  & 12.44\%    & 18.89\% & 30.44\% & 48\% \\ 
Coefficient Error $<$ 10\% & 13.33\%    & 24.44\% & 43.33\% & 59.78\% \\ 

\midrule
\textbf{Standard Error} \\
\cmidrule(lr){1-1}
Standard Error Median Error   & 1.84\%   & 0.00\%  & 5.94\%   & 8.77\% \\
Standard Error Error $<$ 1\%  & 11.58\%  & 22.37\% & 20\%  & 27.63\% \\ 
Standard Error Error $<$ 10\% & 20.53\%  & 28.95\% & 38.42\%  & 46.58\% \\ 

\midrule
\textbf{P-value \& Significant Level} \\
\cmidrule(lr){1-1}
P-value Average Error         & .0385   & .0160   & .0765   & .0371 \\
P-value Median abs Error      & .0001   & 0       & .001   & .0017 \\
P-value Error $<$ 1\%         & 15.79\% & 27.63\% & 43.16\% & 53.42\% \\ 
P-value Error $<$ 10\%        & 24.21\% & 35.53\% & 63.16\% & 79.47\% \\ 
Significant Level Correctness & 25.26\% & 31.58\% & 57.11\% & 73.68\% \\ 
Significant Level Error == 1  & 2.11\%  & 3.95\%   & 7.37\% & 7.89\% \\ 
Significant Level Error == 2  & 0.53\%  & 0        & 3.16\% & 2.63\% \\ 
Significant Level Error == 3  & 1.05\%  & 0.66\%   & 3.42\% & 1.84\% \\ 

\midrule
\textbf{Partial Replication based on Task Type}\\
\cmidrule(lr){1-1}
OLS, Panel OLS                          & 24.35\% & 23.91\% & 43.04\% & 46.09\% \\
Propensity Score Regression             & 3.33\%  & 16.67\% & 56.67\% & 93.33\% \\
Instrument Variable (IV)                & 0       & 25\%    & 20\% & 21.25\% \\
Differences in Differences (DID)        & 0       & 25\%    & 26.67\% & 30\% \\
Regression Discontinuity Designs (RDD)  & 8\%     & 15\%    & 12    & 2\% \\
\cmidrule(lr){1-5}
Data Processing                & 20\%    & 25\%    & 51.74\% & 49.13\%\\
Covariate Adjustment           & 19.31\% & 25\%    & 38.28\% & 42.41\% \\
Fixed Effects                  & 12.5\%  & 16\%    & 32.08\% & 34.58\% \\

\bottomrule
Number of Tasks    & 225 & 180 & 450 & 450
\end{tabular}

\end{adjustbox}
\caption{{\bf . Paper Testset Performance}}
\label{tab:paper_result}
\end{table}

%% file: Bibliography/Bibliography.bib
@ARTICLE{Yoganarasimhan:2024,
  author = {Hema Yoganarasimhan and Irina Iakovetskaia},
  title = {From Feeds to Inboxes: A Comparative Study of Polarization in {Facebook} and Email News Sharing},
  journal = {Management Science},
  year = {2024},
  volume = {70},
  number = {9},
  pages = {6461-6472}
}

@article{Stroube:2024,
author = {Stroube, Bryan K. and Waguespack, David M.},
title = {Status and consensus: Heterogeneity in audience evaluations of female- versus male-lead films},
journal = {Strategic Management Journal},
volume = {45},
number = {5},
pages = {994-1024},
doi = {https://doi.org/10.1002/smj.3575},
url = {https://onlinelibrary.wiley.com/doi/abs/10.1002/smj.3575},
year = {2024}
}

@ARTICLE{Abraham:2024,
  author = {Abraham, J. and Olbert, M. and Vasvari, F.},
  title = {{ESG Disclosures in the Private Equity Industry}},
  journal = {Journal of Accounting Research},
  year = {2024},
  volume = {62},
  number = {5},
  pages = {1611-1660}
}

@misc{Wu:2023,
      title={{BloombergGPT}: {A} Large Language Model for Finance}, 
      author={Shijie Wu and Ozan Irsoy and Steven Lu and Vadim Dabravolski and Mark Dredze and Sebastian Gehrmann and Prabhanjan Kambadur and David Rosenberg and Gideon Mann},
      year={2023},
      url={https://arxiv.org/abs/2303.17564}, 
}

@misc{Yao:2023,
      title={Knowledge Plugins: Enhancing Large Language Models for Domain-Specific Recommendations}, 
      author={Jing Yao and Wei Xu and Jianxun Lian and Xiting Wang and Xiaoyuan Yi and Xing Xie},
      year={2023},
      eprint={2311.10779},
      archivePrefix={arXiv},
      primaryClass={cs.IR},
      url={https://arxiv.org/abs/2311.10779}, 
}

@article{Liu:2025,
title = {Integrating chemistry knowledge in large language models via prompt engineering},
journal = {Synthetic and Systems Biotechnology},
volume = {10},
number = {1},
pages = {23-38},
year = {2025},
issn = {2405-805X},
doi = {https://doi.org/10.1016/j.synbio.2024.07.004},
url = {https://www.sciencedirect.com/science/article/pii/S2405805X24001029},
author = {Hongxuan Liu and Haoyu Yin and Zhiyao Luo and Xiaonan Wang},
}

@article{Singhal:2023,
	author = {Singhal, Karan and Tu, Tao and Gottweis, Juraj and Sayres, Rory and Wulczyn, Ellery and Amin, Mohamed and Hou, Le and Clark, Kevin and Pfohl, Stephen R. and Cole-Lewis, Heather and Neal, Darlene and Rashid, Qazi Mamunur and Schaekermann, Mike and Wang, Amy and Dash, Dev and Chen, Jonathan H. and Shah, Nigam H. and Lachgar, Sami and Mansfield, Philip Andrew and Prakash, Sushant and Green, Bradley and Dominowska, Ewa and Ag{\"u}era y Arcas, Blaise and Toma{\v s}ev, Nenad and Liu, Yun and Wong, Renee and Semturs, Christopher and Mahdavi, S. Sara and Barral, Joelle K. and Webster, Dale R. and Corrado, Greg S. and Matias, Yossi and Azizi, Shekoofeh and Karthikesalingam, Alan and Natarajan, Vivek},
	journal = {Nature Medicine},
	number = {3},
	pages = {943-950},
	title = {Toward expert-level medical question answering with large language models},
	volume = {31},
	year = {2025}}

@article{Anderson:2008,
title = {Safety for whom? {The} effects of light trucks on traffic fatalities},
journal = {Journal of Health Economics},
volume = {27},
number = {4},
pages = {973-989},
year = {2008},
issn = {0167-6296},
doi = {https://doi.org/10.1016/j.jhealeco.2008.02.001},
url = {https://www.sciencedirect.com/science/article/pii/S0167629608000040},
author = {Michael Anderson}
}

@inproceedings{Xu:2024,
title={Do Large Language Models Have Compositional Ability? {An} Investigation into Limitations and Scalability},
author={Zhuoyan Xu and Zhenmei Shi and Yingyu Liang},
booktitle={ICLR 2024 Workshop on Mathematical and Empirical Understanding of Foundation Models},
year={2024},
url={https://openreview.net/forum?id=4XPeF0SbJs}
}

@misc{Ling:2024,
      title={Domain Specialization as the Key to Make Large Language Models Disruptive: A Comprehensive Survey}, 
      author={Chen Ling and Xujiang Zhao and Jiaying Lu and Chengyuan Deng and Can Zheng and Junxiang Wang and Tanmoy Chowdhury and Yun Li and Hejie Cui and Xuchao Zhang and Tianjiao Zhao and Amit Panalkar and Dhagash Mehta and Stefano Pasquali and Wei Cheng and Haoyu Wang and Yanchi Liu and Zhengzhang Chen and Haifeng Chen and Chris White and Quanquan Gu and Jian Pei and Carl Yang and Liang Zhao},
      year={2024},
      eprint={2305.18703},
      archivePrefix={arXiv},
      primaryClass={cs.CL},
      url={https://arxiv.org/abs/2305.18703}, 
}

@ARTICLE{Bursztyn:2022,
  author = {Leonardo Bursztyn and Aakaash Rao and Christopher Roth and David Yanagizawa-Drott},
  title = {Opinions as Facts},
  journal = {The Review of Economic Studies},
  year = {2022},
  volume = {90},
  pages = {1832–1864},
  doi = {https://doi.org/10.1093/restud/rdac065}
}

@ARTICLE{Graeber:2024,
  author = {Thomas Graeber and Christopher Roth and Florian Zimmermann},
  title = {Stories, Statistics, and Memory},
  journal = {The Quarterly Journal of Economics},
  year = {2024},
  volume = {139},
  number = {4},
  pages = {2181–2225},
  doi = {https://doi.org/10.1093/qje/qjae020}
}

@article{Ahrens:2024,
title = {Mind your language: Market responses to central bank speeches},
journal = {Journal of Econometrics},
pages = {105921},
year = {2025},
volume = {249},
issn = {0304-4076},
doi = {https://doi.org/10.1016/j.jeconom.2024.105921},
url = {https://www.sciencedirect.com/science/article/pii/S0304407624002720},
author = {Maximilian Ahrens and Deniz Erdemlioglu and Michael McMahon and Christopher J. Neely and Xiye Yang}
}

@article{Gorodnichenko:2024,
title = {Central bank communication on social media: What, to whom, and how?},
journal = {Journal of Econometrics},
pages = {105869},
year = {2025},
volume = {249},
issn = {0304-4076},
doi = {https://doi.org/10.1016/j.jeconom.2024.105869},
url = {https://www.sciencedirect.com/science/article/pii/S0304407624002148},
author = {Yuriy Gorodnichenko and Tho Pham and Oleksandr Talavera}
}

@article{Curti:2023,
title = {{Let's face it: Quantifying the impact of nonverbal communication in FOMC press conferences}},
journal = {Journal of Monetary Economics},
volume = {139},
pages = {110-126},
year = {2023},
issn = {0304-3932},
doi = {https://doi.org/10.1016/j.jmoneco.2023.06.007},
url = {https://www.sciencedirect.com/science/article/pii/S0304393223000740},
author = {Filippo Curti and Sophia Kazinnik}
}

@article{Gmyrek:2024,
author = {Gmyrek, Paweł and Lutz, Christoph and Newlands, Gemma},
title = {A technological construction of society: Comparing {GPT-4} and human respondents for occupational evaluation in the {UK}},
journal = {British Journal of Industrial Relations},
year = {2024},
volume = {63},
pages = {180–208},
doi = {https://doi.org/10.1111/bjir.12840},
url = {https://onlinelibrary.wiley.com/doi/abs/10.1111/bjir.12840},
}

@article{Meltzer:2024,
author = {Meltzer, Joshua P.},
title = {{The Impact of Foundational {AI} on International Trade, Services and Supply Chains in Asia}},
journal = {Asian Economic Policy Review},
volume = {19},
number = {1},
pages = {129-147},
doi = {https://doi.org/10.1111/aepr.12451},
url = {https://onlinelibrary.wiley.com/doi/abs/10.1111/aepr.12451},
year = {2024}
}

@article{Armstrong:2024,
title = {Measuring firm exposure to government agencies},
journal = {Journal of Accounting and Economics},
pages = {101703},
year = {2024},
issn = {0165-4101},
doi = {https://doi.org/10.1016/j.jacceco.2024.101703},
url = {https://www.sciencedirect.com/science/article/pii/S0165410124000338},
author = {Daphne M. Armstrong and Stephen Glaeser and Jeffrey L. Hoopes}
}

@article{Goli:2024,
author = {Goli, Ali and Singh, Amandeep},
title = {Frontiers: Can Large Language Models Capture Human Preferences?},
journal = {Marketing Science},
volume = {43},
number = {4},
pages = {709-722},
year = {2024},
doi = {10.1287/mksc.2023.0306},
URL = {https://doi.org/10.1287/mksc.2023.0306}
}

@article{Niu:2024,
author = {Niu, Yimeng and Wu, Jing and Jiang, Shenyang and Jiang, Zhibin},
title = {The Bullwhip Effect in Servitized Manufacturers},
journal = {Management Science},
year = {2025},
volume = {71},
number = {1},
pages = {1-20}
}

@article{Hui:2024,
author = {Hui, Xiang and Reshef, Oren and Zhou, Luofeng},
title = {The Short-Term Effects of Generative Artificial Intelligence on Employment: Evidence from an Online Labor Market},
journal = {Organization Science},
volume = {35},
number = {6},
pages = {1977-1989},
year = {2024},
doi = {10.1287/orsc.2023.18441},
URL = {https://doi.org/10.1287/orsc.2023.18441}
}

@article{Noailly:2024,
title = {Heard the news? {Environmental} policy and clean investments},
journal = {Journal of Public Economics},
volume = {238},
pages = {105190},
year = {2024},
issn = {0047-2727},
doi = {https://doi.org/10.1016/j.jpubeco.2024.105190},
url = {https://www.sciencedirect.com/science/article/pii/S0047272724001269},
author = {Joëlle Noailly and Laura Nowzohour and Matthias {van den Heuvel} and Ireneu Pla}
}

@article{Leib:2023,
    author = {Leib, Margarita and Köbis, Nils and Rilke, Rainer Michael and Hagens, Marloes and Irlenbusch, Bernd},
    title = {Corrupted by Algorithms? {How} {AI}-generated and Human-written Advice Shape (Dis)honesty},
    journal = {The Economic Journal},
    volume = {134},
    number = {658},
    pages = {766-784},
    year = {2023},
    month = {09},
    issn = {0013-0133},
    doi = {10.1093/ej/uead056},
    url = {https://doi.org/10.1093/ej/uead056}
}

@article{Balsmeier:2024,
    author = {Balsmeier, Benjamin and Fleming, Lee and Stiebale, Joel and Veihl, Maria},
    title = {The Unintended Impact of {R\&D} Tax Credits on Innovative Search},
    journal = {The Review of Economics and Statistics},
    pages = {forthcoming},
    year = {2024},
    month = {11},
    issn = {0034-6535},
    doi = {10.1162/rest_a_01534},
    url = {https://doi.org/10.1162/rest\_a\_01534}
}

@misc{OpenAI:2024,
      title={{GPT-4} Technical Report}, 
      author={OpenAI},
      year={2023},
      url={https://arxiv.org/abs/2303.08774}, 
}

@article{Brynjolfsson:2024,
      title={Generative {AI} at Work}, 
      author={Erik Brynjolfsson and Danielle Li and Lindsey Raymond},
      year={2025},
      journal = {The Quarterly Journal of Economics},
      volume = {140},
      number = {2},
      pages = {889–942}
}

@misc{Dell:2023,
  title={Navigating the jagged technological frontier: Field experimental evidence of the effects of {AI} on knowledge worker productivity and quality},
  author={Dell'Acqua, Fabrizio and McFowland III, Edward and Mollick, Ethan R and Lifshitz-Assaf, Hila and Kellogg, Katherine and Rajendran, Saran and Krayer, Lisa and Candelon, Fran{\c{c}}ois and Lakhani, Karim R},
  url = {https://ssrn.com/abstract=4573321},
  year={2023}
}

@article{Doshi:2024,
author = {Anil R. Doshi  and Oliver P. Hauser },
title = {Generative {AI} enhances individual creativity but reduces the collective diversity of novel content},
journal = {Science Advances},
volume = {10},
number = {28},
pages = {eadn5290},
year = {2024},
doi = {10.1126/sciadv.adn5290},
URL = {https://www.science.org/doi/abs/10.1126/sciadv.adn5290},

}

@article{Noy:2023,
  title={Experimental evidence on the productivity effects of generative artificial intelligence},
  author={Noy, Shakked and Zhang, Whitney},
  journal={Science},
  volume={381},
  number={6654},
  pages={187--192},
  year={2023},
  publisher={American Association for the Advancement of Science}
}

@article{Chen:2025,
author = {Chen, Yang and Kirshner, Samuel N. and Ovchinnikov, Anton and Andiappan, Meena and Jenkin, Tracy},
title = {A Manager and an {AI} Walk into a Bar: Does {ChatGPT} Make Biased Decisions Like We Do?},
journal = {Manufacturing \& Service Operations Management},
year = {2025},
volume  = {27},
number = {2},
pages = {354-368},
doi = {10.1287/msom.2023.0279},
URL = { https://doi.org/10.1287/msom.2023.0279}
}

@misc{He:2024,
      title={Does Prompt Formatting Have Any Impact on {LLM} Performance?}, 
      author={Jia He and Mukund Rungta and David Koleczek and Arshdeep Sekhon and Franklin X Wang and Sadid Hasan},
      year={2024},
      eprint={2411.10541},
      archivePrefix={arXiv},
      primaryClass={cs.CL},
      url={https://arxiv.org/abs/2411.10541}, 
}

@article{Mueller:2019,
   author = {Mueller-Langer, Frank and Fecher, Benedikt and Harhoff, Dietmar and Wagner, Gert G.},
   title = {Replication studies in economics—{How} many and which papers are chosen for replication, and why?},
   journal = {Research Policy},
   volume = {48},
   number = {1},
   pages = {62-83},
   ISSN = {00487333},
   DOI = {10.1016/j.respol.2018.07.019},
   year = {2019},
   type = {Journal Article}
}

@article{Korinek:2023,
   author = {Korinek, Anton},
   title = {Generative {AI} for Economic Research: Use Cases and Implications for Economists},
   journal = {Journal of Economic Literature},
   volume = {61},
   number = {4},
   pages = {1281-1317},
   ISSN = {0022-0515},
   DOI = {10.1257/jel.20231736},
   year = {2023},
   type = {Journal Article}
}

@article{Li:2024,
author = {Li, Peiyao and Castelo, Noah and Katona, Zsolt and Sarvary, Miklos},
title = {Frontiers: Determining the Validity of Large Language Models for Automated Perceptual Analysis},
journal = {Marketing Science},
volume = {43},
number = {2},
pages = {254-266},
year = {2024},
doi = {10.1287/mksc.2023.0454},
URL = {https://doi.org/10.1287/mksc.2023.0454},
}

@article{Chakraborty:2025,
author = {Chakraborty, Ishita and Chiong, Khai and Dover, Howard and Sudhir, K.},
title = {Can {AI} and {AI}-Hybrids Detect Persuasion Skills? {Salesforce} Hiring with Conversational Video Interviews},
journal = {Marketing Science},
volume = {44},
number = {1},
pages = {30-53},
year = {2025},
doi = {10.1287/mksc.2023.0149},
URL = {https://doi.org/10.1287/mksc.2023.0149},
}

@inproceedings{MetaGPT:2024,
      title={Meta{GPT}: Meta Programming for A Multi-Agent Collaborative Framework},
      author={Hong, Sirui and Zhuge, Mingchen and Jonathan Chen and Xiawu Zheng and Yuheng Cheng and Jinlin Wang and Ceyao Zhang and Zili Wang and Steven Ka Shing Yau and Zijuan Lin and Liyang Zhou and Chenyu Ran and Lingfeng Xiao and Chenglin Wu and J{\"u}rgen Schmidhuber},
      booktitle={The Twelfth International Conference on Learning Representations},
      year={2024},
      url={https://openreview.net/forum?id=VtmBAGCN7o},
}

@misc{datainterpreter:2024,
      title={{Data Interpreter: An LLM Agent For Data Science}}, 
      author={Hong, Sirui and Lin, Yizhang and Bang Liu and Bangbang Liu and Binhao Wu and Ceyao Zhang and Chenxing Wei and Danyang Li and Jiaqi Chen and Jiayi Zhang and Jinlin Wang and Li Zhang and Lingyao Zhang and Min Yang and Mingchen Zhuge and Taicheng Guo and Tuo Zhou and Wei Tao and Xiangru Tang and Xiangtao Lu and Xiawu Zheng and Xinbing Liang and Yaying Fei and Yuheng Cheng and Zhibin Gou and Zongze Xu and Chenglin Wu},
      year={2024},
      eprint={2402.18679},
      archivePrefix={arXiv},
      primaryClass={cs.AI},
      url={https://arxiv.org/abs/2402.18679},
}

@ARTICLE{Almond:2005,
  author = {Almond, D. and Chay, K.Y. and Lee, D.S.},
  title = {The Costs of Low Birth Weight},
  journal = {The Quarterly Journal of Economics},
  year = {2005},
  volume = {120},
  number = {3},
  pages = {1031–1083}
}

@inproceedings{Wang:2020,
	author = {Wang, Alex and Pruksachatkun, Yada and Nangia, Nikita and Singh, Amanpreet and Michael, Julian and Hill, Felix and Levy, Omer and Bowman, Samuel},
	booktitle = {Advances in Neural Information Processing Systems 32},
	title = {{SuperGLUE}: {A} Stickier Benchmark for General-Purpose Language Understanding Systems},
        pages = {3266-3280},
	year = {2019}}

@inproceedings{Larochelle:2008,
  title={Zero-data learning of new tasks.},
  author={Larochelle, Hugo and Erhan, Dumitru and Bengio, Yoshua},
  booktitle={Proceedings of the 23rd National Conference on Artificial Intelligence},
  pages={646 - 651},
  year={2008}
}

@ARTICLE{Lampert:2014,
  author={Lampert, Christoph H. and Nickisch, Hannes and Harmeling, Stefan},
  journal={IEEE Transactions on Pattern Analysis and Machine Intelligence}, 
  title={Attribute-Based Classification for Zero-Shot Visual Object Categorization}, 
  year={2014},
  volume={36},
  number={3},
  pages={453-465},
  doi={10.1109/TPAMI.2013.140}
}

@INPROCEEDINGS{Xia:2024,
  author={Xia, Yuchen and Kim, Jiho and Chen, Yuhan and Ye, Haojie and Kundu, Souvik and Hao, Cong Callie and Talati, Nishil},
  booktitle={2024 IEEE International Symposium on Workload Characterization (IISWC)}, 
  title={Understanding the Performance and Estimating the Cost of {LLM} Fine-Tuning}, 
  year={2024},
  volume={},
  number={},
  pages={210-223},
  doi={10.1109/IISWC63097.2024.00027}
}

@article{Cattaneo:2010,
author = {Cattaneo, M.D.},
title = {Efficient semiparametric estimation of multi-valued treatment effects under ignorability},
journal = {Journal of Econometrics},
volume = {155},
number = {2},
pages = {138-154},
year = {2010},
}

@article{ROSENBAUM:1983,
author = {Rosenbaum, P.R. and Rubin, D.B.},
title = {The central role of the propensity score in observational studies for causal effects},
journal = {Biometrika},
volume = {70},
number = {1},
pages = {41-55},
year = {1983},
}

@article{Card:1988,
author = {Card, D. and Sullivan, D.},
title = {Measuring the Effect of Subsidized Training Programs on Movements In and Out of Employment},
journal = {Econometrica},
volume = {56},
number = {3},
pages = {497-530},
year = {1988},
}

@misc{Ludwig:2025,
      title={Large Language Models: An Applied Econometric Framework}, 
      author={Ludwig, Jens and Mullainathan, Sendhil and Rambachan, Ashesh},
      year={2025},
      eprint={2412.07031},
      archivePrefix={arXiv},
      primaryClass={econ.EM},
      url={https://arxiv.org/abs/2412.07031}, 
}

@misc{Manning:2024,
      title={Automated Social Science: Language Models as Scientist and Subjects}, 
      author={Manning, Benjamin S. and Zhu, Kehang and Horton, John J.},
      year={2024},
      eprint={2404.11794},
      archivePrefix={arXiv},
      primaryClass={econ.GN},
      url={https://arxiv.org/abs/2404.11794}, 
}

@misc{Han:2025,
      title={Mining Causality: {AI}-Assisted Search for Instrumental Variables}, 
      author={Han, Sukjin},
      year={2025},
      eprint={2409.14202},
      archivePrefix={arXiv},
      primaryClass={econ.EM},
      url={https://arxiv.org/abs/2409.14202}, 
}

@article{Liebowitz:1995,
author = {Liebowitz, J.},
title = {Expert Systems: A Short Introduction},
journal = {Engineering Fracture Mechanics},
volume = {50},
number = {5-6},
pages = {601-607},
year = {1995},
}

@book{Tzafestas:1993,
    author = {Tzafestas, S.},
    title = {Expert Systems in Engineering Applications},
    publisher = {Springer Berlin, Heidelberg},
    year = {1993}
}

@article{Tan:2017,
author = {Tan, H},
title = {A brief history and technical review of the expert system research},
journal = {IOP Conference Series: Materials Science and Engineering},
volume = {242},
number = {1},
pages = {012111},
year = {2017},
url = {https://dx.doi.org/10.1088/1757-899X/242/1/012111},
}
